\documentclass[11pt, reqno]{amsart}
\usepackage{graphicx,amsmath,amssymb,epsfig}
\usepackage{amsfonts}

\usepackage{hyperref}

\hypersetup{
    colorlinks=true,
    urlcolor=blue,
    linkcolor=blue,
    citecolor=blue
}

\newtheorem{theorem}{Theorem}
\theoremstyle{plain}

\newtheorem{definition}{Definition}
\newtheorem{example}{Example}

\newtheorem{lemma}{Lemma}

\newtheorem{proposition}{Proposition}
\newtheorem{remark}{Remark}

\numberwithin{equation}{section}

\begin{document}
\title[Comonotonic measures of multivariate risks]{Comonotonic measures of
multivariate risks}
\author{Ivar Ekeland$^\dag$ Alfred Galichon$^\S $ \ \ Marc Henry$^\ddag$ }
\date{First version is October 16, 2007. The present version is of November 1, 2009.
The authors thank two anonymous referees and an associate Editor who helped
improving the paper. The authors are grateful to Guillaume Carlier,
Rose-Anne Dana, Nicole El Karoui, Jean-Charles Rochet, Ludger R\"{u}%
schendorf, and Nizar Touzi, as well as participants to the workshop
\textquotedblleft Multivariate and Dynamic Risk Measures\textquotedblright\
(IHP, Paris, October 2008), the SAS\ seminar at Princeton University, the
Statistics Department Risk seminar at Columbia University for helpful
discussions and comments. Galichon and Ekeland gratefully acknowledge
support from Chaire EDF-Calyon \textquotedblleft Finance and D\'{e}%
veloppement Durable\textquotedblright\ and FiME, Laboratoire de Finance des March\'es de l'Energie (www.fime-lab.org), and Galichon that of Chaire Soci\'{e}%
t\'{e} G\'{e}n\'{e}rale \textquotedblleft Risques
Financiers\textquotedblright\, and Chaire Axa \textquotedblleft Assurance et
Risques Majeurs\textquotedblright. }

\begin{abstract}
We propose a multivariate extension of a well-known characterization by
S.~Kusuoka of regular and coherent risk measures as maximal correlation
functionals. This involves an extension of the notion of comonotonicity to
random vectors through generalized quantile functions. Moreover, we propose
to replace the current law invariance, subadditivity and comonotonicity
axioms by an equivalent property we call \emph{strong coherence} and that we
argue has more natural economic interpretation. Finally, we reformulate the
computation of regular and coherent risk measures as an optimal
transportation problem, for which we provide an algorithm and implementation.
\end{abstract}

\maketitle

\noindent


{\footnotesize \textbf{MSC 2000 subject classification}: 91B06, 91B30, 90C08 %
\vskip50pt }

\newpage

%\newpage

%\setcounter{page}{1}

\section*{Introduction}

The notion of coherent risk measure was proposed by Artzner, Delbaen, Eber
and Heath in \cite{ADEH:99} as a set of axioms to be verified by a
real-valued measure of the riskiness of an exposure. In addition to
monotonicity, positive homogeneity and translation invariance, the proposed
coherency axioms include subadditivity, which is loosely associated with
hedging. Given this interpretation, it is natural to require the risk
measure to be additive on the subsets of risky exposures that are \emph{%
comonotonic}, as this situation corresponds to the worse-case scenario for
the correlation of the risks. In \cite{Kusuoka:2001}, Kusuoka showed the
remarkable result that law invariant coherent risk measures that are also
comonotonic additive are defined by the integral of the quantile function
with respect to a positive measure, a family that includes Expected
Shortfall (also known as Conditional Value at Risk, or Expected Tail Loss).

The main drawback of this formulation is that it does not properly handle
the case when the num\'{e}raires in which the risky payoffs are labeled are
not perfect substitutes. This situation is commonly met in Finance. In a
two-country economy with floating exchange rates, the fact that claims on
payoffs in different currencies are not perfectly substitutable is known as
the \emph{Siegel paradox}; in the study of the term structure of interest
rates, the fact that various maturities are (not) perfect substitutes is
called the (failure of the) \emph{pure expectation hypothesis}. The
technical difficulty impeding a generalization to the case of a multivariate
risk measure is that the traditional definition of comonotonicity relies on
the order in $\mathbb{R}$. When dealing with portfolios of risk that are non
perfectly substituable, as Jouini, Meddeb and Touzi did in \cite{JMT:2004}
for coherent risk measures, and R\"{u}schendorf in \cite{Ruschendorf:2006}
for law invariant convex risk measures, the right notion of multivariate
comonotonicity is not immediately apparent.

The present work circumvents these drawbacks to generalize Kusuoka's result
to multivariate risk portfolios, and proposes a simplifying reformulation of
the axioms with firm decision theoretic foundations. First, we propose an
alternative axiom called \emph{strong coherence}, which is equivalent to the
axioms in \cite{Kusuoka:2001} and which, unlike the latter, extends to the
multivariate setting. We then make use of a variational characterization of
Kusuoka's axioms and representation in order to generalize his results to
the multivariate case. We show that multivariate risk measures that satisfy
strong coherence have the same representation as in \cite%
{Kusuoka:2001}, which we discuss further below.

The work is organized as follows. The first section motivates a new notion
called strong coherence which is shown to be intimately related to existing
risk measures axioms, yet appears to be more natural. The second section
shows how the concept of comonotonic regular risk measures can be extended
to the case of multivariate risks, by introducing a proper generalization of
the notion of comonotonicity and giving a representation theorem. The third
section discusses in depth the relation with Optimal Transportation Theory,
and shows important examples of actual computations.

\subsection*{Notations and conventions}

Let $(\Omega ,{\mathcal{F}},\mathbb{P})$ be a probability space, which is
\emph{standard} in the terminology of \cite{JST:2006}, that is $\mathbb{P}$
is nonatomic and $L^{2}(\Omega ,\mathcal{F},\mathbb{P})$ is separable. Let $%
X:\Omega \rightarrow {\mathbb{R}}^{d}$ be a random vector; we denote the
distribution law of $X$ by ${\mathcal{L}}_{X}$, hence ${\mathcal{L}}_{X}=X\#%
\mathbb{P}$, where $X\#\mathbb{P}:=\mathbb{P}X^{-1}$ denotes the \emph{%
push-forward} of probability measure $\mathbb{P}$ by $X$. The \emph{%
equidistribution class} of $X$ is the set of random vectors with
distribution with respect to $\mathbb{P}$ equal to ${\mathcal{L}}_{X}$
(reference to $\mathbb{P}$ will be implicit unless stated otherwise). As
explained in the appendix, essentially one element in the equidistribution
class of $X$ has the property of being the gradient of a convex function;
this random element is called the (generalized) \emph{quantile function}
associated with the distribution ${\mathcal{L}}_{X}$ and denoted by $Q_{X}$
(in dimension 1, this is the quantile function of distribution $\mathcal{L}%
_{X}$ in the usual sense). We denote by ${\mathcal{M}}({\mathcal{L}},{%
\mathcal{L}}^{\prime })$ the set of probability measures on $\mathbb{R}%
^{d}\times \mathbb{R}^{d}$ with marginals ${\mathcal{L}}$ and ${\mathcal{L}}%
^{\prime }$. We call $L_{d}^{2}(\mathbb{P})$ (abbreviated in $L_{d}^{2}$)
the equivalence class of $\mathcal{F}$-measurable functions $\Omega
\rightarrow \mathbb{R}^{d}$ with a finite second moment modulo $\mathbb{P}$%
-negligible events. We call $\mathcal{P}_{2}(\mathbb{R}^{d})$ the set of
probability on $\mathbb{R}^{d}$ with finite second moment. Finally, for two
elements $X,Y$ of $L_{d}^{2}$, we write $X\sim Y$ to indicate equality in
distribution, that is ${\mathcal{L}}_{X}={\mathcal{L}}_{Y}$. We also write $%
X\sim {\mathcal{L}}_{X}$. Define $\mathrm{c.l.s.c.}(\mathbb{R}^{d})$ as the
class of convex lower semi-continuous functions on $\mathbb{R}^{d}$, and the
\emph{Legendre-Fenchel conjugate} of $V\in \mathrm{c.l.s.c.}(\mathbb{R}^{d})$
as $V^{\ast }(x)=\sup_{y\in \mathbb{R}^{d}}\left[ x\cdot y-V(y)\right] $. In
all that follows, the dot \textquotedblleft $\cdot $\textquotedblright\ will
denote the standard scalar product in $\mathbb{R}^{d}$ or $L_{d}^{2}$. $%
\mathcal{M}_{d}(\mathbb{R})$ denotes the set of $d\times d$ matrices, and $%
\mathcal{O}_{d}(\mathbb{R})$ the orthogonal group in dimension $d$. For $%
M\in \mathcal{M}_{d}(\mathbb{R})$, $M^{T}$ denotes the matrix transpose of $M
$. For a function $V:\mathbb{R}^{d}\rightarrow \mathbb{R}$ differentiable at
$x$, we denote $\nabla V\left( x\right) $ the gradient of $V$ at $x$; this
is the vector $\left( \frac{\partial V\left( x\right) }{\partial x_{1}},...,%
\frac{\partial V\left( x\right) }{\partial x_{d}}\right) \in \mathbb{R}^{d}$%
. When $V$ is twice differentiable at $x$, we denote $D^{2}V$ the Hessian
matrix of $V$ that is the matrix $\left( \frac{\partial ^{2}V\left( x\right)
}{\partial x_{i}\partial x_{j}}\right) _{1\leq i,j\leq d}$. By Aleksandrov's
theorem, a convex function is (Lebesgue-) almost everywhere differentiable
on the interior of its domain (see \cite{Villani:2003}, pp. 58--59), so $%
\nabla V$ and $D^{2}V$ exist almost everywhere. For a functional $\Phi $
defined on a Banach space, we denote $D\Phi $ its Fr\'{e}chet derivative.%
\label{RRK9}

\section{Strong coherence: a natural axiomatic characterization}

\label{section: strongly coherent risk measures} In this section we advocate
a very simple axiomatic setting, called \emph{strong coherence} which will
be shown to be equivalent to the more classical axiomatic framework
described in the next section. We argue that this axiom has more intuitive
appeal than the classical (equivalent) axioms.

\subsection{Motivation: Structure Neutrality}

The regulating instances of the banking industry are confronted with the
problem of imposing rules to the banks to determine the amount of regulatory
capital they should budget to cover their risky exposure. A notable example
of such a rule is the Value-at-Risk, imposed by the Basel II committee, but
a number of competing rules have been proposed. We call $X\in L_{d}^{2}$ the
vector of random losses\footnote{%
In this paper we have chosen to restrict ourselves to the case where risks
are in $L^{2}({\mathbb{R}^{d}})$ for notational convenience, but all results
in the paper carry without difficulty over to the case where the risks are
in in $L^{p}({\mathbb{R}^{d}})$ for $p\in (1,+\infty )$.} of a given bank.
Note that contrary to a convention often adopted in the literature, we chose
to account positively for net losses: $X$ is a vector of effective losses.
Also note that we have supposed that the risk is multivariate, which means
that there are multiple num\'{e}raires, which, depending on the nature of
the problem, can be several assets, several term maturities, or several
non-monetary risks of different nature. We suppose that these multiple num%
\'{e}raires cannot be easily exchanged into one another: \emph{the problem
is intrinsiquely multivariate}. This could be the case if the firm (or the
regulator) is unable or unwilling to define a monetary equivalent for the
various dimensions of its risks. For instance, an oil company is likely to
be unable to estimate a dollar amount to price its environmental risk;
similarly, a pharmaceutical company may be unwilling to give a monetary
estimate for the health hazard its product carry.\label{RRK10}

To a random vector of losses $X$ one associates a number $\varrho(X)$ which measures the intensity of the risk incurred. The unit in which $\varrho$ is to be thought of as some extra currency unity, or alternatively a non-monetary score; it is not assumed to be one of the monetary units associated with the various dimensions of the vector of the risks. This score is used by investors to compare the risks of two companies, or by regulators to set limits to risky exposures for regulated firms.
An important desirable feature of the rule proposed by the regulator is to
avoid regulatory arbitrage. Here, a regulatory arbitrage would be possible
if the firms could split their risk into several different subsidiaries $%
S_{i}$, $i=1,...,N$ with independent legal existence, so that the the
shareholder's economic risk remained the same $X=X_{1}+...+X_{N}$, but such
that the amount of the shareholder's capital which is required to be
budgeted to cover their risk were strictly inferior after the split, namely
such that $\varrho (X)>\varrho (X_{1})+...+\varrho (X_{N})$. To avoid this,
we shall impose the requirement of \emph{subadditivity}, that is
\begin{equation*}
\varrho (X_{1}+...+X_{N})\leq \varrho (X_{1})+...+\varrho (X_{N})
\end{equation*}%
for all possible dependent risk exposures $(X_{1},...,X_{N})\in
(L_{d}^{2})^{N}$. We now argue that the regulator is only interested in the
amount and the intensity of the risk, not in its operational nature: the
capital budgeted should be the same for a contingent loss of 1\% of the
total capital at risk no matter how the loss occurred (whether on the
foreign exchange market, the stock market, the credit market, etc.) This
translates mathematically into the requirement that the regulatory capital
to budget should only depend on the distribution of the risk $X$, that is,
the rule should satisfy the \emph{law invariance} property:

\begin{definition}
A functional $\varrho: L^2\rightarrow\mathbb{R}$ is called \emph{%
law-invariant} if $\varrho(X)=\varrho(Y)$ when $X\sim Y$, where $\sim $
denotes equality in distribution.
\end{definition}

By combining together subadditivity and law invariance, we get the natural
requirement for the capital budgeting rule, that $\varrho (\tilde{X}_{1}+...+%
\tilde{X}_{N})\leq \varrho (X_{1})+...+\varrho (X_{N})$ for all $X$,$\tilde{X%
}$ in $(L_{d}^{2})^{N}$ such that $X_{i}\sim \tilde{X_{i}}$ for all $%
i=1,...,N$. However, in order to prevent giving a premium to conglomerates,
and to avoid imposing an overconservative rule to the regulated firms, one is led to
impose the inequality to be sharp and pose the \emph{structure neutrality}
axiom
\begin{equation*}
\varrho (X_{1})+...+\varrho (X_{N})=\sup_{\tilde{X}_{i}\sim X_{i}}\varrho (%
\tilde{X}_{1}+...+\tilde{X}_{N})
\end{equation*}%
This requirement is notably failed by the Value-at-Risk, which leads to the
fact that \emph{the Value-at-Risk as a capital budgeting rule is not neutral
to the structure of the firm}. This result should be read in the perspective
of the corporate finance literature on the optimal structure of the firm,
originating in the celebrated Modigliani-Miller theorem, according to which
the value of the firm does not depend on the structure of its capital.\label%
{RRK11} This point is explained in detail in \cite{Galichon:2008}, where an
explicit construction is provided. We introduce the axiom of strong
coherence to be satisfied by a measure of the riskiness of a portfolio of
risk exposures (potential losses) $X\in L_{d}^{2}$.

\begin{definition}[Strong coherence]
For $\mu\in \mathcal{P}_{2}(\mathbb{R}^{d})$, a functional $\varrho :L_{d}^{2}\rightarrow \mathbb{R}$ is called a \emph{%
strongly coherent risk measure} if (i) it is convex continuous, and (ii) it
is \emph{structure neutral}: for all $X,Y\in {\mathcal{L}}_{d}^{2}$,
\begin{equation*}
\varrho (X)+\varrho (Y)=\sup \left\{ \varrho (\tilde{X}+\tilde{Y})\,:\;%
\tilde{X}\sim X;\;\tilde{Y}\sim Y\;\right\} .
\end{equation*}
\end{definition}

The convexity axiom can be justified by a risk aversion principle: in
general, one should prefer to diversify risk. The structure neutrality
axiom, being defined as a supremum over all correlation structures, can be
interpreted as a provision against worst-case scenarios, and may be seen as
unduly conservative. However, this axiom is no more conservative than the
set of axioms defining a regular coherent risk measure as we shall see.

As we shall see also, strongly coherent risk measures satisfy all the classical
axioms of coherent risk measures (recalled in definition \ref%
{definition-coherent} below) let alone monotonicity, for which the
multivariate extension is not obvious. In particular, these measures satisfy positive homogeneity and translation invariance. They also satisfy law invariance, which
can be seen by taking $Y=0$ in the definition above.
\label{RRK6}

We now show that strongly coherent risk measures are represented by maximal
correlation functionals with respect to a given random vector or scenario.

\subsection{Characterization of strongly coherent risk measures}

We are now going to show that the strong coherence property essentially
characterizes a class of risk measures known as \emph{maximal correlation
risk measures}, which we shall first recall the definition of.

\subsubsection{Maximal correlation measures}

We first define \emph{maximal correlation risk measures} (in the
terminology of R\"{u}schendorf who introduced them in the multivariate case, see e.g. \cite{Ruschendorf:2006}). These measures will generalize the variational formulation for coherent regular risk measures given in (\ref{equation:
polya}) below.

\begin{definition}[Maximal correlation measures]
A functional $\varrho _{\mu }:L_{d}^{2 }\rightarrow \mathbb{R}$ is called a
\emph{maximal correlation risk measure} with respect to a \emph{baseline
distribution} $\mu$ if for all \ $X\in L_{d}^{2 }$,
\begin{equation*}
\varrho _{\mu }(X):= \sup \left\{ \mathbb{E}[X\cdot \tilde{U}]:\;\tilde{U}%
\sim \mu \right\}.
\end{equation*}
\label{deinition: max-correlation}
\end{definition}

Our notion of maximal correlation risk measure is essentially the
same as R\"{u}schendorf's, with a few minor variants: R\"{u}schendorf
defines his measures on $L_{d}^{\infty }$ instead of $L_{d}^{2}$, and
imposes the extra requirements $U_{i}\geq 0$ and $E_{\mu }\left[ U_{i}\right]
=1$ for $i=1,...,d$, which we do not impose for now.\label{RRK1}

\begin{remark}[Geometric interpretation]
The maximum correlation measure with respect to measure $\mu $ is the
support function of the equidistribution class of $\mu $.
\end{remark}

\begin{example}[Multivariate Expected Shortfall]
An interesting example of univariate risk measure within the class of
maximal correlation risk measures is the expected shortfall, also known as
conditional value at risk. This risk measure can be generalized to the
multivariate setting by defining the $\alpha $-expected shortfall of a risk
exposure $X$ as the maximal correlation measure when the baseline risk $U$
is a Bernoulli random vector (i.e. its distribution ${\mathcal{L}}_{U}$ is
determined by $U=(1/\alpha ,\ldots ,1/\alpha )^{T}$ \label{RRK2}with
probability $\alpha $ and $0$ with probability $1-\alpha $). In such case,
one can easily check that if $\mathcal{L}_{X}$ is absolutely continuous,
then defining $W(x)=max(\sum_{i=1}^{d}x_{i}-c,0)$, with $c$ given by
requirement $Pr(\sum_{i=1}^{d}x_{i}\geq c)=\alpha $, it follows that $W$ is
convex and $\nabla W$ exists $\mathcal{L}_{X}$ almost everywhere and pushes $%
\mathcal{L}_{X}$ to $\mathcal{L}_{U}$ as in proposition~\ref{proposition:
maxcorrel}; therefore the maximal correlation measure is given in this case
by $E\left[ \left( \sum_{i=1}^{d}X_{i}\right) 1\{{\sum_{i=1}^{d}X_{i}\geq c}%
\}\right] $. In other words, the maximum correlation measure in this example
is the (univariate) $\alpha $-expected shortfall for $Y=\sum_{i=1}^{d}X_{i}$%
. \label{example: MES}
\end{example}

\begin{example}
With a more complex baseline risk, other important examples where explicit
or numerical computation is possible include the cases when 1) the baseline
risk and the risk to be measured are both Gaussian, or 2) the baseline risk
is uniform on $\left[ 0,1\right] ^{d}$ and the risk to be measured has a
discrete distribution. Both these cases are treated in detail in Section~4.
\end{example}

Let us first recall the following lemma, which emphasizes the symmetry
between the roles played by the equivalence class of $X$ and $U$ in the
definition above.

\begin{lemma}
For any choice of $U\sim \mu $, with $\mu \in \mathcal{P}_{2}(\mathbb{R}%
^{d}) $, one has
\begin{equation*}
\varrho _{\mu }(X)=\sup \left\{ \mathbb{E}[\tilde{X}\cdot U]:\;\tilde{X}\sim
X\right\} ,
\end{equation*}%
and $U$ is called the \emph{baseline risk} associated with $\varrho _{\mu }$%
. It follows that $\varrho _{\mu }$ is law invariant. \label{lemma: rotation}
\end{lemma}

\begin{proof}
See (2.12) in \cite{Ruschendorf:2006}.
\end{proof}

\subsubsection{Characterization}

We now turn to our first main result, which is a characterization of
strongly coherent risk measures. We first prove a useful intermediate
characterization in proposition \ref%
{proposition:alternative-strong-coherence} below. We shall use Lemma A.4
from \cite{JST:2006}, which we quote here for the reader's convenience.
Denote by $\mathcal{A}$ the set of bimeasurable bijections $\sigma $ from $%
\left( \Omega ,\mathcal{A},\mathbb{P}\right) $ into itself which preserve
the probability, so that $\sigma \#\mathbb{P}=\mathbb{P}$. Recall that $%
\left( \Omega ,\mathcal{F},\mathbb{P}\right) $ was assumed to be a
probability space which does not have atoms, and such that $L^{2}\left(
\Omega ,\mathcal{F},\mathbb{P}\right) $ is separable.

\begin{lemma}
Let $C$ be a norm closed subset of $L^{2}\left( \Omega ,\mathcal{F},\mathbb{P%
}\right) $. Then the following are equivalent:

\begin{enumerate}
\item $C$ is law invariant, that is $X\in C$ and $X\sim Y$ implies that $%
Y\in C$

\item $C$ is transformation invariant, that is for any $X\in C$ and any $%
\sigma \in \mathcal{A}$, we have $X\circ \sigma \in C$
\end{enumerate}
\end{lemma}

As an immediate consequence, we have the following result:

\begin{proposition}
\label{proposition:alternative-strong-coherence} A convex continuous
functional $\varrho :L_{d}^{2}\rightarrow \mathbb{R}$ is a strongly coherent
risk measure if and only if we have:
\begin{equation}
\varrho (X)+\varrho (Y)=\sup \left\{ \varrho (X\circ \sigma +Y\circ \tau
)\,:\;\sigma ,\tau \in \mathcal{A}\right\} .  \label{ce}
\end{equation}
\end{proposition}

\begin{proof}
Clearly $X\circ \sigma \sim X$ and $Y\circ \tau \sim Y $. Hence:%
\begin{equation}
\sup \left\{ \varrho (X\circ \sigma +Y\circ \tau )\,:\;\sigma ,\tau \in
\mathcal{A}\right\} \leq \sup \left\{ \varrho (\tilde{X}+\tilde{Y})\,:\tilde{%
X}\sim X,\ \tilde{Y}\sim Y\right\}  \label{ce1}
\end{equation}

To prove the converse, take any $\varepsilon >0$ and some $X^{\prime }\sim X$
and $Y^{\prime }\sim Y$ such that:%
\begin{equation*}
\varrho (X^{\prime }+Y^{\prime })\,\geq \sup \left\{ \varrho (\tilde{X}+%
\tilde{Y})\,:\tilde{X}\sim X,\ \tilde{Y}\sim Y\right\} -\varepsilon
\end{equation*}%
Consider the set $\left\{ X\circ \sigma :\ \sigma \in \mathcal{A}\right\} $
and denote by $C$ its closure in $L^{2}$. It is obviously transformation
invariant. By the preceding Lemma, it is also law invariant. Since $X\in C$
and $X^{\prime }\sim X$, we must have $X^{\prime }\in C$, meaning that there
exists a sequence $\sigma _{n}\in \mathcal{A}$ with $\left\Vert X\circ
\sigma _{n}-X^{\prime }\right\Vert \longrightarrow 0$. Similarly, there must
exist a sequence $\tau _{n}\in \mathcal{A}$ with $\left\Vert Y\circ \tau
_{n}-Y^{\prime }\right\Vert \longrightarrow 0$. Since $\varrho $ is
continuous, it follows that, for $n$ large enough, we have:%
\begin{eqnarray*}
\sup \left\{ \varrho (X\circ \sigma +Y\circ \tau )\,:\;\sigma ,\tau \in
\mathcal{A}\right\} &\geq &\varrho \left(
X\circ\sigma_{n}+Y\circ\tau_{n}\right) \geq \varrho (X^{\prime }+Y^{\prime
})-\varepsilon \, \\
&\geq &\sup \left\{ \varrho (\tilde{X}+\tilde{Y})\,:\tilde{X}\sim X,\ \tilde{%
Y}\sim Y\right\} -2\varepsilon
\end{eqnarray*}%
and since this holds for any $\varepsilon >0$, the converse of (\ref{ce1})
holds
\end{proof}

We can now state our main result:

\begin{theorem}
Let $\left( \Omega ,\mathcal{F},\mathbb{P}\right) $ be a probability space
which does not have atoms, and such that $L^{2}\left( \Omega ,\mathcal{F},%
\mathbb{P}\right) $ is separable. Let $\varrho $ be a functional defined on $%
{L}_{d}^{2}$. Then the following propositions are equivalent:

\begin{description}
\item[(i)] $\varrho $ is a strongly coherent risk measure;

\item[(ii)] $\varrho $ is a maximal correlation risk measure
\end{description}

\label{theorem: representation}
\end{theorem}
Before we turn to the proof, note that this representation implies immediately that strongly coherent risk measures are in particular positive homogenous and translation invariant, as announced above.
\begin{proof}
We first show (i)$\Rightarrow $(ii). As the proof is quite long, we will
punctuate it with several lemmas.

By the preceding proposition and law invariance, it is enough to prove that:%
\begin{equation}
\varrho \left( X\right) +\varrho \left( Y\right) =\sup_{\sigma \in \mathcal{A%
}}\varrho \left( X+Y\circ \sigma \right)  \label{1}
\end{equation}

Call $\varrho ^{\ast }$ the Legendre transform of $\varrho $ in $L_{d}^{2}$.

\begin{lemma}
$\varrho ^{\ast }$ is law-invariant.
\end{lemma}

\begin{proof}
For $\sigma \in \mathcal{A}$, one has $\varrho ^{\ast }\left( X^{\ast }\circ
\sigma \right) =\sup_{X\in L_{d}^{2}}\left\{ \langle \ X^{\ast }\circ \sigma
,X\rangle -\varrho \left( X\right) \right\} $, so $\varrho ^{\ast }\left(
X^{\ast }\circ \sigma \right) =\sup_{X\in L_{d}^{2}}\left\{ \langle \
X^{\ast },X\circ \sigma ^{-1}\ \rangle -\varrho \left( X\right) \right\}
=\sup_{X\in L_{d}^{2}}\left\{ \langle \ X^{\ast },X\circ \sigma ^{-1}\
\rangle -\varrho \left( X\circ \sigma ^{-1}\right) \right\} =\varrho ^{\ast
}\left( X^{\ast }\right) .$
\end{proof}

\begin{lemma}
\label{lemma: dualinf} If the functions $f_{i},i\in I$, are l.s.c. convex
functions, then%
\begin{equation*}
\left( \sup_{i}f_{i}\right) ^{\ast }=\left( \inf_{i}f_{i}^{\ast }\right)
^{\ast \ast }
\end{equation*}
\end{lemma}

\begin{proof}
For a given l.s.c. convex function $f$, $f\leq \left( \sup_{i}f_{i}\right)
^{\ast }$ is equivalent to $f^{\ast }\geq \sup_{i}f_{i}$, hence to $f\geq
f_{i}$ for all $i$, hence to $f^{\ast }\leq f_{i}^{\ast }$ for all $i$,
hence $f\leq \inf_{i}f_{i}^{\ast }$, hence, as $f$ is l.s.c. convex, to $%
f\leq \left( \inf_{i}f_{i}^{\ast }\right) ^{\ast \ast }$, QED.
\end{proof}

Applying lemma \ref{lemma: dualinf} to the structure neutrality equation,
one has
\begin{eqnarray*}
\varrho ^{\ast }\left( X^{\ast }\right) +\varrho ^{\ast }\left( Y^{\ast
}\right)  &=&\left( \inf_{\sigma \in \mathcal{A}}\sup_{X,Y}\left\{ \langle \
X,X^{\ast }\ \rangle +\langle \ Y,Y^{\ast }\ \rangle -\varrho \left(
X+Y\circ \sigma \right) \right\} \right) ^{\ast \ast } \\
&=&\left( \inf_{\sigma \in \mathcal{A}}\sup_{Y}\left\{ \langle \ Y,Y^{\ast
}\ \rangle +\varrho ^{\ast }\left( X^{\ast }\right) -\langle \ Y\circ \sigma
,X^{\ast }\ \rangle \right\} \right) ^{\ast \ast } \\
&=&\left( \varrho ^{\ast }\left( X^{\ast }\right) +\inf_{\sigma \in \mathcal{%
A}}\sup_{Y}\langle \ Y,Y^{\ast }-X^{\ast }\circ \sigma ^{-1}\ \rangle
\right) ^{\ast \ast }.
\end{eqnarray*}%
The term in $\sup_{Y}\left( ...\right) $ on the right-hand side is $0$ if $%
Y^{\ast }=X^{\ast }\circ \sigma ^{-1}$ and $+\infty $ otherwise. Hence the
previous formula becomes
\begin{equation}
\varrho ^{\ast }\left( X^{\ast }\right) +\varrho ^{\ast }\left( Y^{\ast
}\right) =\varphi ^{\ast \ast }\left( X^{\ast },Y^{\ast }\right)   \label{2}
\end{equation}%
where we have defined
\begin{equation}
\varphi \left( X^{\ast },Y^{\ast }\right) =%
\begin{array}{c}
\varrho ^{\ast }\left( X^{\ast }\right) \text{ if }X^{\ast }\sim Y^{\ast }
\\
+\infty \text{ otherwise.}%
\end{array}
\label{3}
\end{equation}

Now suppose $\varphi \left( X^{\ast },Y^{\ast }\right) <\infty $, hence that
$\varrho ^{\ast }\left( X^{\ast }\right) =\varrho ^{\ast }\left( Y^{\ast
}\right) <\infty $ and $X^{\ast }\sim Y^{\ast }$. As $\varphi \geq \varphi
^{\ast \ast }$, it follows that $\varrho ^{\ast }\left( X^{\ast }\right)
\geq \varrho ^{\ast }\left( X^{\ast }\right) +\varrho ^{\ast }\left( Y^{\ast
}\right) $ hence $\varrho ^{\ast }\left( Y^{\ast }\right) =\varrho ^{\ast
}\left( X^{\ast }\right) \leq 0$, and $\varphi \left( X^{\ast },Y^{\ast
}\right) \leq 0$.

Suppose $\varphi \left( X^{\ast },Y^{\ast }\right) <\infty $ and $\varphi
\left( X^{\ast },Y^{\ast }\right) -\varphi ^{\ast \ast }\left( X^{\ast
},Y^{\ast }\right) <\varepsilon $. Replacing in (\ref{2}), one finds that
\begin{equation*}
0\leq -\varrho ^{\ast }\left( X^{\ast }\right) =-\varrho ^{\ast }\left(
Y^{\ast }\right) \leq \varepsilon
\end{equation*}

\begin{lemma}
$\varphi ^{\ast \ast }$ is valued into $\left\{ 0,+\infty \right\} $.
\end{lemma}

\begin{proof}
As $\varphi ^{\ast }=\varphi ^{\ast \ast \ast }$, one has
\begin{eqnarray*}
\varphi ^{\ast }\left( X,Y\right) &=&\sup_{\left( X^{\ast },Y^{\ast }\right)
}\left\{ \langle \ X,X^{\ast }\ \rangle +\langle \ Y,Y^{\ast }\ \rangle
-\varphi ^{\ast \ast }\left( X^{\ast },Y^{\ast }\right) \right\} \\
&=&\sup_{\left( X^{\ast },Y^{\ast }\right) }\left\{ \langle \ X,X^{\ast }\
\rangle +\langle \ Y,Y^{\ast }\ \rangle -\varphi \left( X^{\ast },Y^{\ast
}\right) \right\} .
\end{eqnarray*}%
Taking a maximizing sequence $\left( X_{n}^{\ast },Y_{n}^{\ast }\right) $ in
the latter expression, one has necessarily $\varphi \left( X_{n}^{\ast
},Y_{n}^{\ast }\right) -\varphi ^{\ast \ast }\left( X_{n}^{\ast
},Y_{n}^{\ast }\right) \longrightarrow 0$. From the previous remark, $%
\varrho ^{\ast }\left( X_{n}^{\ast }\right) =\varrho ^{\ast }\left(
Y_{n}^{\ast }\right) \longrightarrow 0$, hence $\varphi \left( X_{n}^{\ast
},Y_{n}^{\ast }\right) \longrightarrow 0$. Therefore%
\begin{equation*}
\varphi ^{\ast }\left( X,Y\right) =\sup_{(X^{\ast },Y^{\ast }):~\varphi
(X^{\ast },Y^{\ast })=0}\left\{ \langle \ X,X^{\ast }\ \rangle +\langle \
Y,Y^{\ast }\ \rangle \right\}
\end{equation*}%
which is clearly positively homogeneous of degree $1$. Its Legendre
transform $\varphi ^{\ast \ast }$ can therefore only take values $0$ and $%
+\infty $, QED.
\end{proof}

Therefore, there is a closed convex set $K$ such that $\varphi ^{\ast \ast }$
is the indicator function of $K$, that is%
\begin{equation}
\varphi ^{\ast \ast }\left( X^{\ast },Y^{\ast }\right) =%
\begin{array}{c}
0\text{ if }\left( X^{\ast },Y^{\ast }\right) \in K \\
+\infty \text{ otherwise}%
\end{array}
\label{5}
\end{equation}%
and condition (\ref{2})\ implies that%
\begin{equation}
\varrho ^{\ast }\left( X^{\ast }\right) +\varrho ^{\ast }\left( Y^{\ast
}\right) =%
\begin{array}{c}
0\text{ if }\left( X^{\ast },Y^{\ast }\right) \in K \\
+\infty \text{ otherwise}%
\end{array}
\label{4}
\end{equation}

Note that if $\varrho ^{\ast }\left( X^{\ast }\right) <\infty $, then $%
\varphi \left( X^{\ast },Y^{\ast }\right) =\varrho ^{\ast }\left( X^{\ast
}\right) $ for all $Y^{\ast }\sim X^{\ast }$, and then $\varphi ^{\ast \ast
}\left( X^{\ast },Y^{\ast }\right) \leq \varphi \left( X^{\ast },Y^{\ast
}\right) <\infty $. This implies that $\varphi ^{\ast \ast }\left( X^{\ast
},Y^{\ast }\right) =0$, hence that $\varrho ^{\ast }\left( X^{\ast }\right)
=0$. Therefore $\varrho ^{\ast }$ is also an indicator function: there
exists a closed convex set $C$ such that%
\begin{equation}
\varrho ^{\ast }\left( X^{\ast }\right) =%
\begin{array}{c}
0\text{ if }X^{\ast }\in C \\
+\infty \text{ otherwise}%
\end{array}
\label{8}
\end{equation}

By comparison of (\ref{4}) and (\ref{8}), one finds that%
\begin{equation*}
K=C\times C
\end{equation*}

By duality, (\ref{8}) becomes%
\begin{eqnarray}
\varrho \left( X\right) &=&\sup_{X^{\ast }\in C}\langle \ X^{\ast },X\
\rangle  \label{7} \\
C &=&\left\{ X^{\ast }\ |\ \varrho ^{\ast }\left( X^{\ast }\right) =0\right\}
\notag
\end{eqnarray}

Condition (\ref{3}) then implies that $\varphi $ is an indicator function:\
there exists a set $K_{0}$ (in general, neither a closed nor a convex set)
such that%
\begin{equation*}
\varphi \left( X^{\ast },Y^{\ast }\right) =%
\begin{array}{c}
0\text{ if }\left( X^{\ast },Y^{\ast }\right) \in K_{0} \\
+\infty \text{ otherwise}%
\end{array}%
\end{equation*}

By comparison with formulas (\ref{3}) and (\ref{5}), one finds that
\begin{eqnarray*}
\left( X^{\ast },Y^{\ast }\right) &\in &K_{0}\Longleftrightarrow X^{\ast
}\in C,\ Y^{\ast }\in C\text{ and }X^{\ast }\sim Y^{\ast } \\
K &=&\overline{co}\text{ }K_{0}
\end{eqnarray*}

\begin{lemma}
Denote by $\mathcal{E}\left( C\right) $ the set of strongly exposed points
of $C$, and $\overline{K_{0}}$ the closure of $K_{0}$ for the norm topology
in $L^{2}\times L^{2}$. Then
\begin{equation*}
\mathcal{E}\left( C\right) \times \mathcal{E}\left( C\right) \subset
\overline{K_{0}}
\end{equation*}
\end{lemma}

\begin{proof}
Recall (cf. \cite{FabulousFab}) that $X^{\ast }$ is \textit{strongly exposed}
in $C$ if there is a continuous linear form $X$ such that any maximizing
sequence for $X$ in $C$ converges strongly to $X^{\ast }:$
\begin{equation*}
\left.
\begin{array}{c}
X_{n}^{\ast }\in C \\
\langle \ X,X_{n}^{\ast }\ \rangle \longrightarrow \sup_{C}\langle \
X,X^{\ast }\ \rangle
\end{array}%
\right\} \text{ }\Longrightarrow \left\Vert X_{n}^{\ast }-X^{\ast
}\right\Vert \longrightarrow 0
\end{equation*}%
For $\varepsilon >0$, denote by $T_{C}\left( X,\varepsilon \right) $ the set
of $Y^{\ast }\in C$ such that $\sup_{Z^{\ast }\in C}\langle \ X,Z^{\ast }\
\rangle -\langle \ X,Y^{\ast }\ \rangle \leq \varepsilon $. Then $X^{\ast
}\in C$ is strongly exposed by $X$ if and only if $\sup_{Z^{\ast }\in
C}\langle \ X,Z^{\ast }\ \rangle =\langle \ X,X^{\ast }\ \rangle $ and $%
\delta \left[ T_{C}\left( X,\varepsilon \right) \right] $ tends to $0$ when $%
\varepsilon \longrightarrow 0$, where $\delta $ denotes the diameter,
\begin{equation*}
\delta \left[ T_{C}\left( X,\varepsilon \right) \right] :=\sup \left\{
\left\Vert X_{1}^{\ast }-X_{2}^{\ast }\right\Vert \ |\ X_{1}^{\ast }\in
T_{C}\left( X,\varepsilon \right) ,\ X_{2}^{\ast }\in T_{C}\left(
X,\varepsilon \right) \right\} .
\end{equation*}%
Going back to the problem, it is clear that if $X^{\ast }$ and $Y^{\ast }$
are strongly exposed in $C$, then $\left( X^{\ast },Y^{\ast }\right) $ is
strongly exposed in $C\times C$:
\begin{equation*}
\mathcal{E}\left( C\right) \times \mathcal{E}\left( C\right) \subset
\mathcal{E}\left( C\times C\right) =\mathcal{E}\left( K\right)
\end{equation*}

We claim that every strongly exposed point of $K$ necessarily belongs to $%
\overline{K_{0}}$ (the closure is still the norm closure). Indeed, suppose
there exists $\left( X_{1}^{\ast },X_{2}^{\ast }\right) \in \mathcal{E}%
\left( K\right) $ such that $\left( X_{1}^{\ast },X_{2}^{\ast }\right)
\notin \overline{K_{0}}$. Then there exists $\varepsilon >0$ such that $%
K_{0}\cap B\left( \left( X_{1}^{\ast },X_{2}^{\ast }\right) ,\varepsilon
\right) =\varnothing $, where $B\left( \left( X_{1}^{\ast },X_{2}^{\ast
}\right) ,\varepsilon \right) $ is the ball of center $\left( X_{1}^{\ast
},X_{2}^{\ast }\right) $ and radius $\varepsilon >0$. As $\left( X_{1}^{\ast
},X_{2}^{\ast }\right) $ is strongly exposed, there exists a linear form $%
\left( X_{1},X_{2}\right) $ strongly exposing it, and one can choose $\eta >0
$ small enough to ensure $\delta \left[ T_{K}\left( \left(
X_{1},X_{2}\right) ,\eta \right) \right] <\varepsilon $. Since $T_{K}\left(
\left( X_{1},X_{2}\right) ,\eta \right) \,$\ contains $\left( X_{1}^{\ast
},X_{2}^{\ast }\right) $, one concludes that $K_{0}\cap T_{K}\left( \left(
X_{1},X_{2}\right) ,\eta \right) =\varnothing $, thus
\begin{equation*}
K_{0}\subset \left\{ \left( Y_{1}^{\ast },Y_{2}^{\ast }\right) \in K\ |\
\langle \ X_{1},Y_{1}^{\ast }\ \rangle +\langle \ X_{2},Y_{2}^{\ast }\
\rangle \geq \ \langle \ X_{1},X_{1}^{\ast }\ \rangle +\langle \
X_{2},X_{2}^{\ast }\ \rangle +\eta \right\}
\end{equation*}%
But the right-hand side is a closed convex set, so by taking the closed
convex hull of the left-hand side, one gets
\begin{equation*}
\overline{co}\left( K_{0}\right) \subset \left\{ \left( Y_{1}^{\ast
},Y_{2}^{\ast }\right) \in K\ |\ \langle \ X_{1},Y_{1}^{\ast }\ \rangle
+\langle \ X_{2},Y_{2}^{\ast }\ \rangle \geq \ \langle \ X_{1},X_{1}^{\ast
}\ \rangle +\langle \ X_{2},X_{2}^{\ast }\ \rangle +\eta \right\}
\end{equation*}%
and taking $\left( Y_{1}^{\ast },Y_{2}^{\ast }\right) =\left( X_{1}^{\ast
},X_{2}^{\ast }\right) \in K$ leads to a contradiction.

Therefore $\mathcal{E}\left( K\right) \subset \overline{K_{0}}$, and one has
$\mathcal{E}\left( C\right) \times \mathcal{E}\left( C\right) \subset
\mathcal{E}\left( K\right) \subset \overline{K_{0}}$, QED.
\end{proof}

By a celebrated theorem of Bishop and Phelps (see again \cite{FabulousFab}),
there is a dense subset $H$ of $L^{2}$ (in fact, a dense $G_{\delta }$) such
that, for every $X\in H$, the maximum of $\langle \ X^{\ast },X\ \rangle $
for $X^{\ast }\in C$ is attained at a strongly exposed point. Going back to (%
\ref{7}), take some $X\in H$, and let $X^{\ast }\in C$ be such that%
\begin{equation*}
\varrho \left( X\right) =\langle \ X^{\ast },X\ \rangle
\end{equation*}

with $X^{\ast }\in \mathcal{E}\left( C\right) $. Now take another $Y\in H$,
and another point $Y^{\ast }\in \mathcal{E}\left( C\right) $ such that $%
\varrho \left( Y\right) =\langle Y^{\ast },Y\rangle $. One has $\left(
X^{\ast },Y^{\ast }\right) \in \mathcal{E}\left( C\right) \times \mathcal{E}%
\left( C\right) $, and it results from the previous lemma that $\left(
X^{\ast },Y^{\ast }\right) \in \overline{K_{0}}$. This implies the existence
of a sequence $\left( X_{n}^{\ast },Y_{n}^{\ast }\right) \in K_{0}$ such
that $\left( X_{n}^{\ast },Y_{n}^{\ast }\right) $ converges to $\left(
X^{\ast },Y^{\ast }\right) $ in norm. By the definition of $K_{0}$, one
should have $X_{n}^{\ast }\sim Y_{n}^{\ast }$, that is $Y_{n}^{\ast
}=X_{n}^{\ast }\circ \sigma _{n}$ for $\sigma _{n}\in \mathcal{A}$. Hence, $%
\varrho \left( Y\right) =\langle \ Y^{\ast },Y\ \rangle =\lim_{n}\langle \
Y_{n}^{\ast },Y\ \rangle =\lim_{n}\langle \ X_{n}^{\ast }\circ \sigma
_{n},Y\ \rangle =\lim_{n}\langle X_{n}^{\ast },Y\circ \sigma
_{n}^{-1}\rangle $. But by the Cauchy-Schwartz inequality,
\begin{equation*}
\left\vert \langle \ X_{n}^{\ast },Y\circ \sigma _{n}^{-1}\ \rangle -\langle
\ X^{\ast },Y\circ \sigma _{n}^{-1}\ \rangle \right\vert \leq \left\Vert
Y\circ \sigma _{n}^{-1}\right\Vert _{2}\left\Vert X_{n}^{\ast }-X^{\ast
}\right\Vert _{2},
\end{equation*}%
which tends to 0 as $\left\Vert Y\circ \sigma _{n}^{-1}\right\Vert
_{2}=\left\Vert Y\right\Vert _{2}$. Therefore, $\varrho \left( Y\right)
=\lim_{n}\langle \ X_{n}^{\ast },Y\circ \sigma _{n}^{-1}\ \rangle
=\lim_{n}\langle \ X^{\ast },Y\circ \sigma _{n}^{-1}\ \rangle \leq \sup_{%
\tilde{Y}\sim Y}\langle \ X^{\ast },\tilde{Y}\ \rangle $. But one has also $%
\varrho \left( Y\right) =\sup_{Y^{\ast }\in C}\langle \ Y^{\ast },Y\ \rangle
\geq \sup_{\sigma \in \mathcal{A}}\langle \ X^{\ast }\circ \sigma ,Y\
\rangle =\sup_{\sigma \in \mathcal{A}}\langle \ X^{\ast },Y\circ \sigma \
\rangle \geq \sup_{\tilde{Y}\sim Y}\langle \ X^{\ast },\tilde{Y}\ \rangle $,
therefore
\begin{equation*}
\varrho \left( Y\right) =\sup_{\tilde{Y}\sim Y}\langle \ X^{\ast },\tilde{Y}%
\ \rangle \ \ \ \forall Y\in \Omega
\end{equation*}%
The functions $\rho \left( Y\right) $ and:\
\begin{equation*}
\sup_{\tilde{Y}\sim Y}\langle \ X^{\ast },\tilde{Y}\ \rangle =\sup_{\tilde{X}%
\sim Y}\langle \ \tilde{X}^{\ast },Y\ \rangle
\end{equation*}%
are both convex, finite and lsc on $L^{2}$, and hence continuous.\ Since
they coincide on a dense subset, they coincide everywhere.\ This proves the
direct implication (i)$\Rightarrow $(ii) of the theorem.

We now turn to the converse. Let $\varrho _{\mu }$ be a maximal correlation
risk measure with respect to baseline measure $\mu $. Then $\varrho _{\mu }$
is clearly convex. Take $X$ and $Y$ in $L_{d}^{2}$. By proposition \ref%
{proposition: maxcorrel} in the Appendix, there exist two convex functions $%
\phi _{1}$ and $\phi _{2}$ such that for $U\sim \mu $, one has $\nabla \phi
_{1}(U)\sim X$ and $\nabla \phi _{2}(U)\sim Y$, and $\varrho _{\mu
}(X)=E[U\cdot \nabla \phi _{1}(U)]$, $\varrho _{\mu }(Y)=E[U\cdot \nabla
\phi _{2}(U)]$. Thus $\varrho _{\mu }(X)+\varrho _{\mu }(Y)=E[U\cdot (\nabla
\phi _{1}(U)+\nabla \phi _{2}(U))]$, but for all $\tilde{U}\sim U$, $%
E[U\cdot (\nabla \phi _{1}(U)+\nabla \phi _{2}(U))]\geq E[\tilde{U}\cdot
(\nabla \phi _{1}(U)+\nabla \phi _{2}(U))]$, hence $\varrho _{\mu
}(X)+\varrho _{\mu }(Y)=\sup \left\{ \varrho (\tilde{X}+\tilde{Y})\,:\;%
\tilde{X}\sim X,\tilde{Y}\sim Y\right\} $. Thus $\varrho _{\mu }$ is
strongly coherent, which completes the proof of Theorem \ref{theorem:
representation}. \hskip4pt
\end{proof}

\section{A multivariate generalization of Kusuoka's theorem}

\label{section: strong coherence and comonotonicity}

In this section we recall the existing axiomatization leading to the
representation result of Kusuoka in \cite{Kusuoka:2001}, where risk measures
for univariate risks that are subadditive, law invariant and comonotonic
additive are represented by maximal correlation functionals. We then propose
a way to generalize these axioms to the case where risk measures deal with
multivariate risks, by showing how to generalize the only problematic axiom,
namely comonotonic additivity. We then give a representation result which
extends Kusuoka's to the multivariate case.

\subsection{Coherent and regular risk measures}

To describe the existing axiomatic framework, we first recall the following
definitions valid in the univariate case, from \cite{ADEH:99}, and existing results.

\begin{definition}[Coherent; Convex risk measures]
\label{definition-coherent}A functional $\varrho :L^{2}\rightarrow \mathbb{R}
$ is called a \emph{coherent risk measure} if it satisfies the following
four properties (MON), (TI), (CO) and (PH) as follows:

\begin{itemize}
\item Monotonicity (MON): $X\leq Y\Rightarrow\varrho(X)\leq\varrho(Y)$

\item Translation invariance (TI): $\varrho(X+m)=\varrho(X)+m\varrho(1)$

\item Convexity (CO): $\varrho(\lambda X+(1-\lambda)
Y)\leq\lambda\varrho(X)+(1-\lambda)\varrho(Y)$ for all $\lambda\in [0,1]$.

\item Positive homogeneity (PH): $\varrho(\lambda X)=\lambda\varrho(X)$ for
all $\lambda\geq0$.
\end{itemize}

A functional which only satisfies (MON), (TI) and (CO) is called a \emph{%
convex risk measure}.
\end{definition}

Even though these definitions are mostly standard, note that since we have
considered risk measures associated with random vectors of potential \emph{%
losses}, the definition of monotonicity takes an non decreasing form, unlike
the definition in most of the literature on coherent risk measures. Compared
to the traditional presentation in the literature, the expression of
translation invariance is adapted to take into account the fact that we did
not impose the scaling convention $\varrho (1)=1$.\label{RRK3} Also note (as
we have a multivariate generalization in mind) that, let alone monotonicity
(which we shall discuss separately below), all these axioms admit a
straightforward generalization to the case of risks $X\in L_{d}^{2}$. The
expression for (CO) and (PH) will remain unchanged; for (TI) the natural
extension to dimension $d$ will be given in (\ref{TIdimd}) below.\label{RRK4}

A representation of coherent risk measures was given in the original work of
\cite{ADEH:99}, whereas representation of convex risk measures was proposed
in \cite{FS:2004}. These were extended to the multivariate setting by
Jouini, Meddeb and Touzi in \cite{JMT:2004} who characterize coherent
acceptance sets, i.e. sets in $\mathbb{R}^n$ that cancel the risk associated
with an $\mathbb{R}^d$ valued random vector, and consider aggregation
issues, and Burgert and R\"uschendorf in \cite{BR:2006} who characterize
convex real valued measures for multivariate risks, and R\"uschendorf in
\cite{Ruschendorf:2006}, who characterizes those of the latter that are law
invariant, and proposes maximal correlation risk measures as an example. The
idea of introducing a variational characterization of comonotonic additivity
as well as the generalization of Kusuoka's axiomatic approach it allows
constitute the essential novelties of this section.

\textbf{Regularity. }In the case of univariate risks, \emph{comonotonic
additivity} is used in addition to law invariance to define \emph{regular
risk measures} (see \cite{FS:2004}, sect. 4.7):

\begin{definition}[Comonotonicity; Regularity]
Two random variables $X$ and $Y$ are \emph{comonotonic} (or synonymously,
\emph{comonotone})\label{RRK7} if there exits a random variable $U$ and two
increasing functions $\phi $ and $\psi $ such that $X=\phi (U)$ and $Y=\psi
(U)$ hold almost surely.

A functional $\varrho: L^2\rightarrow\mathbb{R}$ is called a \emph{regular
risk measure} if it satisfies:

\begin{itemize}
\item Law invariance (LI), and

\item Comonotonic additivity (CA): $\varrho(X+Y)=\varrho(X)+\varrho(Y)$ when
$X,Y$ are comonotonic.
\end{itemize}
\end{definition}

Note that comonotonic additivity implies translation invariance, as any
random variable is comonotonic with the constant. Informally speaking, law
invariance suggests that the risk measure is a functional of the quantile
function $F_{X}^{-1}(t)=\inf \{x:F_{X}(x)\geq t\}$ associated with the
distribution. Positive homogeneity and comonotonic additivity together
suggest that this representation is linear $\varrho (X):=\int_{0}^{1}\phi
(t)F_{X}^{-1}(t)dt$. Finally, subadditivity suggests that the weights $\phi
(t)$ are increasing with respect to $t$. Precisely Kusuoka has shown the
following in \cite{Kusuoka:2001}, Theorem~7:

\begin{proposition}[Kusuoka]
A coherent risk measure $\varrho $ is regular if and only if for some
increasing and nonnegative function $\phi $ on $[0,1]$, we have
\begin{equation*}
\varrho (X):=\int_{0}^{1}\phi (t)F_{X}^{-1}(t)dt,
\end{equation*}%
where $F_{X}$ denotes the cumulative distribution functions of the random
variable $X$, and its generalized inverse $F_{X}^{-1}(t)=\inf
\{x:F_{X}(x)\geq t\}$ is the associated quantile function. \label%
{proposition: Kusuoka}
\end{proposition}

\textbf{Variational characterization. }By the Hardy-Littlewood-P\'{o}lya
inequality shown in lemma~11 of \cite{Kusuoka:2001}, we can write a
variational expression for coherent regular risk measures:
\begin{equation}
\int_{0}^{1}\phi (t)F_{X}^{-1}(t)dt=\max \left\{ \mathbb{E}[X\tilde{U}]:\;%
\tilde{U}\sim \mu \right\} .  \label{equation: polya}
\end{equation}%
where $\mu $ if the probability distribution of $\phi $, and the maximum is
taken over the equidistribution class of $\mu $. The reader is referred to
\cite{Dana:2005} and \cite{CL} for a nice treatment of this variational
problem and the dual representation of Schur convex functions in the
univariate case\label{RRK8}. As we shall see, variational
characterization \ref{equation: polya} will be key when generalizing to the multivariate setting.

\subsection{A multivariate notion of comonotonicity}

We now turn to an extension of the concept of comonotonicity. Note first
that a valid definition of comonotonicity in dimension one is the following:
two random variables $X$ and $Y$ are comonotonic if and only if one can
construct almost surely $Y=T_Y(U)$ and $X=T_X(U)$ for some third random
variable $U$, and $T_X$, $T_Y$ non decreasing functions. In other words, $X$
and $Y$ are comonotonic whenever there is a random variable $U$ such that $%
\mathbb{E}[UX]=\max\left\{\mathbb{E}[X\tilde{U}]:\; \tilde{U}\sim U\right\}$
and $\mathbb{E}[UY]=\max\left\{\mathbb{E}[Y\tilde{U}]:\;\tilde{U}\sim
U\right\}$. This variational characterization will be the basis for our
generalized notion of comonotonicity.

To simplify our exposition in the remainder of the paper, we shall make the
following assumption:

\textbf{Assumption. }\emph{In the remainder of the paper, we shall assume
that the baseline distribution of risk $\mu$ is absolutely continuous with
respect to Lebesgue measure.}

\begin{definition}[$\protect\mu$-comonotonicity]
Let $\mu$ be a probability measure on $\mathbb{R}^d$ that is absolutely
continuous. Two random vectors $X$ and $Y$ in $L_d^2$ are called $\mu $%
-comonotonic if for some random vector $U\sim \mu $, we have
\begin{eqnarray*}
U&\in &argmax_{\tilde{U}}\left\{ \mathbb{E}[X \cdot \tilde{U}],\;\tilde{U}%
\sim \mu \right\} \text{, and} \\
U& \in & argmax_{\tilde{U}}\left\{ \mathbb{E}[Y \cdot \tilde{U}],\;\tilde{U}%
\sim \mu \right\} .
\end{eqnarray*}%
\label{definition: comonotonicity}
\end{definition}

In particular, every random vector $X$ is $\mu $-comonotonic with constant
vectors $Y=y$. Note that the geometric interpretation of this definition is
that $X$ and $Y$ are $\mu $-comonotonic if and only if they have the same $%
L^{2}$ projection on the equidistribution class of $\mu $. We next give a
few useful lemmas. We start with a result securing the existence of a $\mu $%
-comonotonic pair with given marginals.

\begin{lemma}
Let $\mu$ be a probability measure on $\mathbb{R}^d$ that is absolutely
continuous. Then given two probability distributions $P$ and $Q$ in $%
\mathcal{P}_2(\mathbb{R}^d)$, there exists a pair $(X,Y)$ in $(L^2_d)^2$
such that $X\sim P$, $Y\sim Q$, and $X$ and $Y$ are $\mu$-comonotonic. \label%
{lemma: comonotonic-repres}
\end{lemma}

\begin{proof}
By Brenier's theorem (Proposition \ref{proposition: maxcorrel} in the
Appendix), there exists $U\sim \mu$ and two convex functions $\phi_1$ and $%
\phi_2$ such that $X=\nabla \phi_1(U) \sim P$ and $Y=\nabla \phi_2(U) \sim Q$%
. Then $X$ and $Y$ are $\mu$-comonotonic.
\end{proof}

We then provide a useful characterization of $\mu$-comonotonicity.

\begin{lemma}
Let $\mu$ be probability measure on $\mathbb{R}^d$ that is absolutely
continuous. Then two random vectors $X$ and $Y$ in $L_d^2$ are $\mu $%
-comonotonic if
\begin{eqnarray*}
\varrho_\mu(X+Y) = \varrho_\mu(X) + \varrho_\mu(Y)
\end{eqnarray*}%
where $\varrho_\mu(X):= \sup \left\{ \mathbb{E}[X\cdot \tilde{U}]:\;\tilde{U}%
\sim \mu \right\}$ is the maximal correlation risk measure, defined in
Definition \ref{deinition: max-correlation} above. \label{lemma:
comonotonic-additive}
\end{lemma}

\begin{proof}
There exists $U\sim\mu$ such that $\varrho_\mu(X+Y)=\mathbb{E}[(X+Y)\cdot U]$%
. We have $\mathbb{E}[(X+Y) \cdot U] = \mathbb{E}[X \cdot U]+\mathbb{E}[Y
\cdot U]$, and both inequalities $\mathbb{E}[X \cdot U] \leq \varrho_\mu(X) $
and $\mathbb{E}[Y \cdot U] \leq \varrho_\mu(Y)$ hold, thus $\mathbb{E}[X
\cdot U]+\mathbb{E}[Y \cdot U ] \leq \varrho_\mu(X) + \varrho_\mu(Y) $ with
equality if and only both inequalities above are actually equalities, which
is the equivalence needed.
\end{proof}

This lemma implies in particular that maximal correlation functionals with baseline measure $\mu$ are $\mu$-comonotone additive. Thus combining with Theorem \ref{theorem: representation}, this establishes that strongly coherent risk measures are $\mu$-comonotone additive for some $\mu$.

We next show that \emph{in dimension 1}, the notion of $\mu$-comonotonicity
is equivalent to the classical notion of comonotonicity, regardless of the
choice of $\mu$ (provided it is absolutely continuous).

\begin{lemma}
In dimension $d=1$, let $\mu$ be probability measure on $\mathbb{R}^d$ that
is absolutely continuous. Then $X$ and $Y$ are $\mu$-comonotonic if and only
if they are comonotonic in the classical sense, that is, if and only if
there exists a random variable $Z$ and two non decreasing functions $f$ and $%
g$ such that $X=f(Z)$ and $Y=g(Z)$ holds almost surely.
\end{lemma}

\begin{proof}
Suppose that $X$ and $Y$ are $\mu$-comonotonic. Then there is a $U\sim \mu$
such that $U \in argmax_{\tilde{U}}\left\{ \mathbb{E}[X \tilde{U}],\;\tilde{U%
}\sim \mu \right\}$ and $U \in argmax_{\tilde{U}}\left\{ \mathbb{E}[Y \tilde{%
U}],\;\tilde{U}\sim \mu \right\}$. This implies in particular the existence
of two increasing functions $f$ and $g$ such that $X=f(U)$ and $Y=g(U)$
holds almost surely. Hence $X$ and $Y$ are comonotonic in the classical
sense. Conversely, suppose that $X$ and $Y$ are comonotonic in the classical
sense. There exists a random variable $Z$ and two increasing functions $f$
and $g$ such that $X=f(Z)$ and $Y=g(Z)$ holds almost surely. Let $F_Z$ be
the cumulative distribution function of $Z$, and $F_\mu$ the one associated
with $\mu$. Defining $U=F^{-1}_\mu\left( F_Z (Z) \right)$, one has $U\sim
\mu $, and denoting $\varphi = f \circ F^{-1}_\mu \circ F_Z$ and $\phi = g
\circ F^{-1}_\mu \circ F_Z$, one has $X=\varphi(U)$ and $Y=\phi(U)$. Thus $X$
and $Y$ are $\mu$-comonotonic.
\end{proof}

In dimension one, one recovers the classical notion of comonotonicity
regardless of the choice of $\mu$ as shown in the previous lemma. However,
in dimension greater than one, the comonotonicity relation crucially depends
on the baseline distribution $\mu $, unlike in dimension one. The following
lemma makes this precise.

\begin{lemma}
Let $\mu$ and $\nu$ be probability measures on $\mathbb{R}^d$ that is
absolutely continuous. Then: \newline
- In dimension $d=1$, $\mu$-comonotonicity always implies $\nu$%
-comonotonicity. \newline
- In dimension $d \geq 2$, $\mu$-comonotonicity implies $\nu$-comonotonicity
if and only if $\nu=T\#\mu$ for some location-scale transform $T(u) =
\lambda u + u_0$ where $\lambda >0$ and $u_0\in{\mathbb{R}}^d$. In other
words, comonotonicity is an invariant of the location-scale family
transformation classes. \label{lemma: comonotonic-related}
\end{lemma}

\begin{proof}
In dimension one, all the notions of $\mu$-comonotonicity coincide with the
classical notion of comonotonicity, as remarked above. Let $d\geq 2$, and
suppose that $\mu$-comonotonicity implies $\nu$-comonotonicity. Consider $U
\sim \mu$, and let $\phi$ be the convex function (defined up to an additive
constant) such that $\nabla \phi \# \nu = \mu$. Then there exists a random
vector $V\sim\nu$ such that $U = \nabla \phi (V) $ almost surely. Consider
some arbitrary symmetric positive endomorphism $\Sigma$ acting on $\mathbb{R}%
^d$. Then the map $u\to \Sigma(u)$ is the gradient of a convex function
(namely the associated quadratic form $u\to \frac 1 2 \left< u, \Sigma(u)
\right>$), therefore the random vectors $U$ and $\Sigma (U)$ are $\mu$%
-comonotonic. By hypothesis, it follows that $U$ and $\Sigma (U)$ are also $%
\nu$-comonotonic, hence there exists a convex function $\zeta$ such that $%
\Sigma (U) = \nabla \zeta (V)$ holds almost surely. Therefore, the equality $%
\Sigma \circ \nabla \phi (v) = \nabla \zeta (v)$ holds for almost every $v$.
By differentiating twice (which can be done almost everywhere, by
Aleksandrov's theorem), we get that $\Sigma \circ D^2\phi (v) = D^2 \zeta (v)$
hence $\Sigma \circ D^2\phi$ is almost everywhere a symmetric endomorphism.
This being true regardless of the choice of $\Sigma$, it follows that the
matrix of $D^2\phi$ in any orthonormal basis of $\mathbb{R}^d$ is almost
everywhere a diagonal matrix, hence there exists a real valued map $%
\lambda(u)$ such that $D^2\phi (u) = \lambda(u) u$, with $\lambda(u)>0$. But
this implies $\partial_{u_i}\partial_{u_j} \phi (u)= 0$ for $i\neq j$ and $%
\partial^2_{u_i}\phi (u)= \lambda(u)$ for all $i$. Therefore, $%
\partial_{u_j} \lambda(u) = \partial_{u_j}\partial^2_{u_i}\phi (u) = 0$.
Hence $\lambda(u)=\lambda $ a strictly positive constant. It follows that $%
\nabla \phi (u) = \lambda u + u_0$, QED. The converse holds trivially.\hskip%
4pt
\end{proof}

\begin{remark}
A close inspection of the proof of this lemma reveals that the essential
reason of the discrepancy between dimension one and higher is the simple
fact that the general linear matrix group $\mathcal{G}l_{d}(\mathbb{R})$ is
Abelian if and only if $d=1$.
\end{remark}

We can now define a concept which generalizes comonotonic additivity to the
multidimensional setting.

\begin{definition}[ $\protect\mu$-comonotonic additivity; $\protect\mu$%
-regularity]
A functional $\varrho: L_d^2\rightarrow\mathbb{R}$ is called a \emph{$\mu$%
-regular risk measure} if it satisfies:

\begin{itemize}
\item Law invariance (LI), and

\item $\mu $-comonotonic additivity ($\mu $-CA): $\varrho (X+Y)=\varrho
(X)+\varrho (Y)$ when $X,Y$ are $\mu $-comonotonic.
\end{itemize}
\end{definition}

As every random vector is comonotonic with constants, it implies that a $\mu
$-comonotonic additive functional $\varrho $ is in particular translation
invariant in the following multivariate sense \label{RRK5}
\begin{equation}
\varrho \left( X+my\right) =\varrho \left( X\right) +m\varrho \left(
y\right) \text{ for all }m\in \mathbb{R}\text{\ and }y\in \mathbb{R}^{d}.
\label{TIdimd}
\end{equation}

\subsection{A multivariate extension of Kusuoka's theorem}

We now show that maximal correlation is equivalent to the combination of
subadditivity, law invariance, $\mu$-comonotonic additivity and positive
homogeneity. Further, the probability measure $\mu$ involved in the
definition of comonotonic additivity shall be precisely related to the one
which is taken as a baseline scenario of the maximal correlation measure.

We have seen above (lemma \ref{lemma: comonotonic-additive}) that maximal
correlation risk measures defined with respect to a distribution $\mu $ are $%
\mu$-comonotonic additive. When the measure is also law invariant and
coherent, we shall see that the converse holds true, and this constitutes
our second main result, which is a multivariate extension of Kusuoka's
theorem. Note that while Kusuoka's theorem was stated using the axioms of
subadditivity and positive homogeneity in addition to others, we only need
the weaker axiom of convexity in addition to the same others.

\begin{theorem}
Let $\varrho$ be a l.s.c. risk measure on ${L}^2_d$ with the properties of convexity (CO), and $\mu$-regularity, that is law invariance (LI) and $\mu$-comonotonic additivity ($\mu$-CA). Then $%
\varrho$ is strongly coherent. Equivalently, $\varrho$ is a maximal correlation risk measure, namely there exists $\nu \in
\mathcal{P}_2(\mathbb{R}^d)$ such that $\varrho=\varrho_\nu$, where $%
\varrho_\nu$ is a maximal correlation measure with respect to baseline
scenario $\nu$, and $\mu$ and $\nu$ are related by location-scale
transformation, that is $\nu=T\#\mu$ where $T(u) = \lambda u + u_0$ with $\lambda >0$ and $%
u_0\in{\mathbb{R}}^d$. \label{theorem: representation2}
\end{theorem}

\begin{proof}
Combining the convexity and law invariance axioms imply $\varrho(\tilde{X}+%
\tilde{Y}) \leq \varrho(X) + \varrho(Y)$ for all $X,Y,\tilde{X},\tilde{Y}$
in $L_{d}^{2}$, thus $\varrho (X)+\varrho (Y)\geq \sup \left\{ \varrho (%
\tilde{X}+\tilde{Y})\,:\;\tilde{X}\sim X;\;\tilde{Y}\sim Y\;\right\}$. But
by Lemma \ref{lemma: comonotonic-repres}, there exists a $\mu$-comonotonic
pair $(X,Y)$. By $\mu$-comonotonic additivity, one has $\varrho
(X)+\varrho(Y)=\varrho(X+Y)$, therefore the previous inequality is actually
an equality, and
\begin{equation*}
\varrho (X)+\varrho (Y)= \sup \left\{ \varrho (\tilde{X}+\tilde{Y})\,:\;%
\tilde{X}\sim X;\;\tilde{Y}\sim Y\;\right\}
\end{equation*}
therefore $\varrho$ is strongly coherent. By Theorem \ref{theorem:
representation}, it results that there exists $\nu \in \mathcal{P}_2(\mathbb{%
R}^d)$ such that $\varrho=\varrho_\nu$. But by the comonotonic additivity of
$\varrho$ and lemma~\ref{lemma: comonotonic-additive}, any two vectors $X$
and $Y$ which are $\mu$-comonotonic are also $\nu$-comonotonic. By lemma~\ref%
{lemma: comonotonic-related}, this implies that there is a location-scale
map $T$ such that $\nu=T\#\mu$, so that the result follows. \hskip4pt
\end{proof}

Because it allows a natural generalization of well-known univariate results,
this theorem makes a strong point in arguing that our notion of comonotonic
additivity is the right one when considering multivariate risks.

\subsection{Extending monotonicity}

We extend the concept of monotonicity with reference to a partial order  $%
\preceq$ defined on $\mathbb{R}^{d}$ in the following way:

\begin{definition}[$\preceq $-monotonicity]
A functional $\varrho :L^{2}\rightarrow \mathbb{R}$ is said to be \emph{$%
\preceq $-monotone} if it satisfies:\newline
($\preceq $-MON): $X\preceq Y$ almost surely $\Rightarrow \varrho (X)\leq
\varrho (Y)$.
\end{definition}

We have the following result:

\begin{proposition}
Let $\varrho_\mu$ be the maximal correlation risk measure with respect to
baseline distribution $\mu$. Let $(Supp~\mu)^0$ be the polar cone of the
support of $\mu$. For a cone $C\subset {\mathbb{R}}^d$, denote $\preceq _{C}$
the partial order in ${\mathbb{R}}^d$ induced by $C$, namely $x \preceq _{C}
y$ if and only if $y-x \in C$. Then $\varrho_\mu$ is monotone with respect
to $\preceq _{C}$ if and only if $C\subset - (Supp~\mu)^0$. \label%
{proposition: monotonicity}
\end{proposition}

\begin{proof}
If $X$ and $U$ are $\mu$-comonotonic, then $D\varrho_X(Z)=E[U\cdot Z]$, but
the property that $E[U \cdot Z] \geq 0$ for all $Z$ almost surely included
in $C$ is equivalent to $C \subset -(Supp~\mu)^0$.\hskip4pt
\end{proof}

Note that in dimension $d=1$, with $C= {\mathbb{R}}_+$, one recovers the
usual notion of monotonicity. In higher dimension, we get in particular that
if $\mu$ is supported in ${\mathbb{R}}_+^d$, then $\varrho_\mu$ is monotone
with respect to the strong order of ${\mathbb{R}}^d$. Finally, note also
that the concept of monotonicity proposed here is a somewhat weak one, as it
deals only with almost sure domination between $X$ and $Y$. A stronger
concept of monotonicity would involve stochastic ordering of $X$ and $Y$; we
do not pursue this approach here.

\section{Numerical computation}

In this section, we show explicit examples of computation of the maximal
correlation risk measure. We start by the Gaussian case, where closed-form
formulas are available. To handle more general cases we shall show that the
problem may be thought of as an auction mechanism, an intuition we shall
develop and use to derive an efficient computational algorithm.

\subsection{Gaussian risks}

We now consider the case where the baseline risk $U$ is Gaussian with
distribution $\mu =N(0,\Sigma _{U})$, with $\Sigma _{U}$ a positive definite
matrix of size $d$, and we study the restriction of $\varrho _{\mu }$ to the
class of Gaussian risks.

Note (cf. \cite{RR:98} I, Ex. 3.2.12) that the linear map $u\rightarrow
A_{X}u$ where
\begin{equation*}
A_{X}=\Sigma _{U}^{-1/2}(\Sigma _{U}^{1/2}\Sigma _{X}\Sigma
_{U}^{1/2})^{1/2}\Sigma _{U}^{-1/2},
\end{equation*}%
sends the probability measure $N(0,\Sigma _{U})$ to the probability measure $%
N(0,\Sigma _{X})$; further, $A_{X}$ is positive semidefinite, so this map is
the gradient of convex function $u\rightarrow \frac{1}{2}u^{\prime }A_{X}u$.
Hence we have the following straightforward matrix formulation of
comonotonicity.

\begin{lemma}
Consider two Gaussian vectors $X\sim N(0,\Sigma _{X})$ and $Y\sim N(0,\Sigma
_{Y})$ with $\Sigma _{X}$ and $\Sigma _{Y}$ invertible. Then $X$ and $Y$ are
$\mu $-comonotonic if and only if
\begin{equation}  \label{eqn: gauss-max-corr}
E[XY^T]= \Sigma_{U}^{-1/2}(\Sigma_{U}^{1/2}\Sigma_{X}\Sigma_{U}^{1/2})
^{1/2}(\Sigma_{U}^{1/2}\Sigma_{Y}\Sigma_{U}^{1/2}) ^{1/2} \Sigma_{U}^{-1/2}.
\end{equation}
In particular, in the case $\mu =N(0,I_{d})$, $X$ and $Y$ are $\mu $%
-comonotonic if and only if $E[XY^T]=\Sigma _{X}^{1/2}\Sigma _{Y}^{1/2}$. %
\label{lemma: gaussian-comon}
\end{lemma}

\begin{proof}
If $X$ and $Y$ are $\mu$-comonotonic, then there exists $U\sim N(0,
\Sigma_U) $ such that $X=A_X U$ and $Y=A_Y U$, and the result follows.
Conversely, if equality (\ref{eqn: gauss-max-corr}) holds, then denoting $%
U=A_X^{-1}X$ and $V=A_Y^{-1}Y$, we get that 1) $U\sim N(0,\Sigma_U)$ and $%
V\sim N(0,\Sigma_U)$, and 2) $E[UV^T]=A_X^{-1}E[XY^T]A_Y^{-1}=\Sigma_U$,
therefore by the Cauchy-Schwartz inequality, $U=V$ almost surely. Thus $X$
and $Y$ are $\mu$-comonotonic. \hskip4pt
\end{proof}

We now derive the value of correlation risk measures at Gaussian risks.
Still by \cite{RR:98} I, Ex. 3.2.12, we have immediately:

\begin{proposition}
When the baseline risk $U$ is Gaussian with distribution $\mu =N(0,\Sigma_U)$%
, we have for a Gaussian vector $X\sim N(0,\Sigma_X )$:
\begin{equation*}
\varrho _{\mu }(X)=tr \left[ \left( \Sigma_U^{1/2} \Sigma_X \Sigma_U^{1/2}
\right)^{1/2} \right].
\end{equation*}
In particular, in the case $\mu =N(0,I_{d})$, $\varrho _{\mu}$ is the trace
norm: $\varrho _{\mu}(X)=tr \left[ \Sigma_X^{1/2} \right]$. \label%
{proposition: gaussian-maxcorr}
\end{proposition}

\begin{proof}
One has $\varrho_\mu(X)=\max\{\mathbb{E}[\tilde{X}\cdot U]; \tilde{X}\sim
X\} =\mathbb{E}\left[A_X U U^T\right]$, thus because of the previous
results, $\varrho_\mu(X)= \mathbb{E}\left[U^T\Sigma_{U}^{-1/2}(%
\Sigma_{U}^{1/2}\Sigma_{X}\Sigma_{U}^{1/2}) ^{1/2}\Sigma_{U}^{-1/2}U\right]=%
\mathrm{tr}\left((\Sigma_{U}^{1/2}\Sigma_{X}\Sigma_{U}^{1/2}) ^{1/2}\right)$.%
\hskip4pt
\end{proof}

In dimension 2, we have the formula $tr\left( \sqrt{S}\right) =\sqrt{%
tr\left( S\right) +2\sqrt{\det S}}$, so we get a closed form expression:

\begin{example}
When $d=2$, and $\mu =N(0,I_{2})$, we have for $\Sigma_X =%
\begin{pmatrix}
\sigma _{1}^{2} & \varrho \sigma _{1}\sigma _{2} \\
\varrho \sigma _{1}\sigma _{2} & \sigma _{2}^{2}%
\end{pmatrix}%
$ the following expression $\varrho _{\mu }(X)=\sqrt{\sigma _{1}^{2}+\sigma
_{2}^{2}+2\sigma _{1}\sigma _{2}\sqrt{1-\varrho ^{2}}}$.
\end{example}

\subsection{Kantorovich duality and Walras auction}

We now see how optimal transportation duality permits the computation of
maximal correlation risk measures. More precisely, we shall see that the
algorithm we shall propose to compute numerically the maximal correlation
risk measures is to be thought of intuitively as a \emph{Walrasian auction},
as we shall explain. We refer to \cite{RR:98} and \cite{Villani:2003} for
overviews of the theory and applications of optimal transportation,
including recent results. Consider a baseline distribution $\mu$, and recall
the expression for the maximal correlation risk measure $\varrho_{\mu}(X)$
of a random vector $X\in\mathbb{R}^d$: $\varrho(X)= \sup\left\{\mathbb{E}[ X
\cdot \tilde{U}]:\; \tilde{U}\sim \mu\right\}$. This problem is the problem
of computing the maximal transportation cost of mass distribution $\mu$ to
mass distribution $\mathcal{L}_X$ with cost of transportation $c(u,x)=u\cdot
x$.

The problem has a dual expression according to Monge-Kantorovich duality (or
duality of optimal transportation). We have (theorem~2.9 page 60 of \cite%
{Villani:2003}):
\begin{equation}
\varrho_{\mu}(X)=\min_{V\in\mathrm{c.l.s.c.}(\mathbb{R}^d)} \left(\int V
d\mu+\int V^\ast d\mathcal{L}_X \right).  \label{equation: MK duality}
\end{equation}
The function $V$ that achieves the minimum in (\ref{equation: MK duality})
exists by theorem~\ref{theorem: representation}(iii) and when $\mathcal{L}_X$
is absolutely continuous, one has $\nabla V^*(X)\sim \mu$ and $%
\varrho_\mu(X)= \mathbb{E}[ X \cdot \nabla V^*(X)] $. In the sequel we shall
make the law invariance of $\varrho_{\mu}$ and the symmetry between the
roles played by the distributions of $X$ and $U$ explicit in the notation by
writing
\begin{equation*}
\varrho_{\mu}(\mathcal{L}_X) :=\varrho(\mu,\mathcal{L}_X):= \varrho_{\mu}(X).
\end{equation*}

\subsubsection{Law-invariant, convex risk measures}

Following \cite{Ruschendorf:2006}, theorem~2.3, the maximum correlation risk
measures are the building blocks of more general convex risk measures. One
has the following result, which was proven by R\"{u}schendorf in the cited
paper.

\begin{proposition}
Let $\varrho$ be a convex measure. Then $\varrho$ is law invariant if and
only if there exists a penalty function $\alpha$ such that
\begin{equation*}
\varrho(X) = \sup_{\mu \in {\mathcal{P}_2(\mathbb{R}^d)}} \varrho_{\mu}(X) -
\alpha(\mu).
\end{equation*}
Furthermore, $\alpha(\mu)$ can be chosen as $\alpha(\mu) = \sup \{
\varrho_\mu(X) : X \in L^2_d ,~\varrho(X) \leq 0 \} $. \label{proposition:
law-inv-convex}
\end{proposition}

\subsubsection{Dual representations of the risk measure.}

The following lemma provides an expression of the conjugate of the maximal
correlation risk measure.

\begin{lemma}
For $W:\mathbb{R}^{d}\rightarrow \mathbb{R}$ convex and lower
semicontinuous, one has
\begin{equation*}
\sup_{P\in \mathcal{P}_{2}(\mathbb{R}^{d})}\left\{ \varrho _{\mu }(P)+\int
WdP\right\} =\int (-W)^{\ast }d\mu .
\end{equation*}%
\label{lemma: Fenchel}
\end{lemma}

\begin{proof}
One has $\int (-W)^* d\mu = \int \sup_y \left\{ u\cdot y + W(y) \right\}
d\mu(u)$, thus $\int (-W)^* d\mu = \sup_{\tau(\cdot)}\int u\cdot \tau(u) +
W(\tau(u)) d\mu(u)$ where the supremum is over all measurable maps $\tau:%
\mathbb{R} \to \mathbb{R}$. Grouping by equidistribution class, one has
\begin{eqnarray*}
\int (-W)^* d\mu &=& \sup_P \left[ \sup_{\tau \# \mu = P}\int u\cdot \tau(u)
d\mu(u) + \int W dP\right] \\
&=& \sup_P \left\{ \varrho_\mu(P) + \int W dP \right\}.
\end{eqnarray*}
\end{proof}

\subsubsection{General equilibrium interpretation}

We now consider then $\varrho(\mu,\mathcal{L}_X)$ for two probability
distributions on $\mathbb{R}^d$, and we interpret $\mu$ as a distribution of
consumers (e.g. insurees) and $\mathcal{L}_X$ as a distribution of goods
(e.g. insurance contracts) in an economy. Consumer with characteristics $u$
derives utility from the consumption of good with attributes $x$ equal to
the interaction $u\cdot x$ of consumer characteristics and good attributes.
Consumer $u$ maximizes utility $u\cdot x$ of consuming good $x$ minus the
\textbf{price} $V^\ast(x)$ of the good. Hence his \textbf{indirect utility}
is $\sup_{x\in\mathbb{R}^d}\left[u\cdot x-V^\ast(x)\right]%
=V^{\ast\ast}(u)=V(u)$. According to equation~(\ref{equation: MK duality}),
the \textbf{total surplus} in the economy $\mathbb{E}[X\cdot U]$ is
maximized for the pair $(V,V^\ast)$ of convex lower semi-continuous
functions on $\mathbb{R}^d$ that minimizes
\begin{equation*}
\Phi(V):=\int V d\mu + \int V^\ast d\mathcal{L}_X.
\end{equation*}
The functional $\Phi$ is convex and its Fr\'echet derivative, when it
exists, is interpreted as the \textbf{excess supply} in the economy, with
value at $h$ equal to $D\Phi(h)=\int h\;d(\mu-\nu_V)$, where $\nu_V :=\nabla
V^* \# \mathcal{L}_X$. Indeed, the convexity of the map $V \to \Phi(V)$
follows from the identity established above in lemma~\ref{lemma: Fenchel}
\begin{equation*}
\Phi(V) = \sup_{\nu \in \mathcal{P}_2(\mathbb{R})^d} \left\{ \varrho(
\mathcal{L}_X ,\nu ) + \int V d\left(\mu -\nu \right) \right\},
\end{equation*}
thus this map is the supremum of functionals that are linear in $V$. The
supremum is attained for $\nu=\nu_V$, hence it follows that $D\Phi_V(h)=\int
h\;d(\mu-\nu_V)$.

Hence, excess supply is zero when the indirect utility $V$ and the prices $%
V^\ast$ are such that $\nu_V=\mu$. With our economic interpretation above,
this can be seen as a \textbf{Walrasian welfare theorem}, where the total
surplus is maximized by the set of prices that equates excess supply to zero.

This general equilibrium interpretation of maximal correlation risk measures
extends to the method of computation of the latter through a gradient
algorithm to minimize the convex functional $\Phi$. This algorithm can be
interpreted as a \textbf{Walrasian t\^atonnement} that adjusts prices to
reduce excess supply $D\Phi_V$. This algorithm is described in more detail
and implemented fully in the case of discretely distributed risks below.

\subsection{Discrete risks}

We now consider the restriction $\varrho _{\mu }$ to the class of risks
whose distribution is discrete. We have in mind in particular the empirical
distribution of a sample of recorded data of the realization of the risk.
The procedure we shall now describe consists in the computation of the
generalized quantile of the discrete distribution, which opens the way for
econometric analysis of maximal correlation risk measures.

\subsubsection{Representation}

Let $X\sim P_{n}$, where $P_{n}=\sum_{k=1}^{n}\pi _{k}\delta _{Y_{k}}$ is a
discrete distribution supported by $\left\{ Y_{1},...,Y_{n}\right\} $, $n$
distinct points in $\mathbb{R}^{d}$. For instance if $P_{n}$ is the
empirical measure of the sample $\left\{ Y_{1},...,Y_{n}\right\} $, then $%
\pi _{k}=1/n$. We are looking for $\varphi :[0,1]^{d}\rightarrow \mathbb{R}%
^{d}$ such that:

(i) for (almost) all $u\in [0,1]^{d}$, $\varphi \left( u\right) \in
\left\{ Y_{1},...,Y_{n}\right\} $

(ii) for all $k\in \left\{ 1,...,n\right\} $, $\mu \left( \varphi
^{-1}\left\{ Y_{k}\right\} \right) =\pi _{k}$ ie. $\varphi $ pushes forward $%
\mu $ to $P_{n}$

(iii) $\varphi = \nabla V$, where $V: {\mathbb{R}}^{d} \to {\mathbb{R}}$ is
a convex function.

It follows from the Monge-Kantorovich duality that there exist weights $%
\left( w_{1},...,w_{n}\right) \in \mathbb{R}^{n}$, such that $V(u)=w^{\ast
}\left( u\right) :=\max_{k}\left\{ \left\langle u,Y_{k}\right\rangle
-w_{k}\right\} $ is the solution. Introduce the functional $\Phi _{\mu }:%
\mathbb{R}^{n}\rightarrow \mathbb{R}$, $\Phi _{\mu }\left( w\right) =\int
w^{\ast }\left( u\right) d\mu \left( u\right) $. The numerical
implementation of the method is based on the following result:

\begin{proposition}
There exist unique (up to an additive constant) weights $w_{1},...,w_{n}$
such that for $w ^{\ast }\left( u\right) =\max_{k}\left\{ \left\langle
u,Y_{k}\right\rangle -w_{k}\right\} $, the gradient map $\varphi =\nabla w
^{\ast }$ satisfies (i), (ii) and (iii) above. The function $w\rightarrow
\Phi _{\mu }\left( w\right) +\sum_{k=1}^{n}\pi _{k}w_{k}$ is convex, and
reaches its minimum at $w=\left( w_{1},...,w_{n}\right) $ defined above. %
\label{proposition: cas-discret}
\end{proposition}

\begin{proof}
By the Knott-Smith optimality criterion (theorem 2.12(i) page 66 of \cite%
{Villani:2003}), there exists a convex function $w$ on the set $%
\{Y_{1},\ldots ,Y_{n}\}$ such that the optimal pair in (\ref{equation: MK
duality}) is $(w,V)$, where $V$ is the Legendre-Fenchel conjugate of $w$,
i.e. the function $V(u)=\sup_{x\in \{Y_{1},\ldots ,Y_{n}\}}\left( u\cdot
x-w(x)\right) =\max_{k}\left( u\cdot Y_{k}-w_{k}\right) $, where $%
w_{k}=w(Y_{k})$ for each $k=1,\ldots ,n$. Note that the subdifferential $%
\partial V$ is a singleton except at the boundaries of the sets $\mathcal{U}%
_{k}=\left\{ u:\arg \max_{i}\left\{ \left\langle u,Y_{i}\right\rangle
-w_{i}\right\} =k\right\} $, so $\nabla V$ is defined $\mathcal{L}_{U}$
almost everywhere. Since for all $k$, and all $u\in \mathcal{U}_{k}$, $%
Y_{k}\in \partial V(u)$, $\nabla V$ satisfies (i). Finally, by Brenier's
Theorem (theorem 2.12(ii) page 66 of \cite{Villani:2003}), $\nabla V$ pushes
$\mathcal{L}_{U}$ forward to $P_{n}$, hence it also satisfies (iii). The
function $\Phi _{\mu }:w\rightarrow \int w^{\ast }\left( u\right) d\mu
\left( u\right) $ is convex, which follows from the equality%
\begin{equation*}
\int w^{\ast }\left( u\right) d\mu \left( u\right) =\max_{\sigma \left(
.\right) }\int \left\langle u,Y_{\sigma \left( u\right) }\right\rangle
-w_{\sigma \left( u\right) }d\mu \left( u\right)
\end{equation*}%
where the maximum is taken over all measurable functions $\sigma :\mathbb{R}%
^{d}\rightarrow \left\{ 1,...,n\right\} $. \hskip4pt
\end{proof}

\subsubsection{The T\^atonnement Algorithm}

The problem is therefore to minimize the convex function $w\rightarrow \Phi
_{\mu ,\pi }\left( w\right) =\Phi _{\mu }\left( w\right) +\sum_{k=1}^{n}\pi
_{k}w_{k}$, which can be done using a gradient approach. To the best of our
knowledge, the idea of using the Monge-Kantorovich duality to compute the
weights using a gradient algorithm should be credited to F. Aurenhammer and
his coauthors. See \cite{AHA:98} and also \cite{RU:2000}. However, by the
economic interpretation seen above, the algorithm's dynamics is the
time-discretization of a ``t\^atonnement process,'' as first imagined by
L\'eon Walras (1874) and formalized by Paul Samuelson (1947) (see \cite%
{Samuelson:47}). Hence to emphasize the economic interpretation, we shall
refer to the algorithm as ``T\^atonnement Algorithm''.

\textbf{The Algorithm. }Initialize the prices $w^0=0$. At each step $m$,
compute $\Phi _{\mu ,\pi }\left( w^m\right)$ and the excess demand $\nabla
\Phi _{\mu ,\pi }\left( w^m\right)$. For a well chosen elasticity parameter $%
\epsilon^m$, update the prices proportionally to excess demand
\begin{equation*}
w^{m+1} = w^m + \epsilon^m \nabla \Phi _{\mu ,\pi }\left( w^m\right)
\end{equation*}
Go to next step, or terminate the algorithm when the excess demand becomes
smaller than a prescribed level. \hskip4pt$\square $

This algorithm requires the evaluation of the function and its gradient. For
this we shall need to compute in turns, for each $k$: 1) $\mathcal{U}%
_{k}=\left\{ u:\arg \max_{i}\left\{ \left\langle u,Y_{i}\right\rangle
-w_{i}\right\} =k\right\} $; 2) $p_{k}=\mu \left( \mathcal{U}_{k}\right) $;
and 3) $u_{k}$ the barycenter of $\left( \mathcal{U}_{k},\mu \right) $ (that
is $u_{k}=\mu (\mathcal{U}_{k})^{-1}\int_{\mathcal{U}_{k}}zd\mu (z)$.) Then
we get the value of $\Phi _{\mu ,\pi }\left( w\right) $: $\Phi _{\mu ,\pi
}\left( w\right) =\sum \left( \left\langle u_{k},Y_{k}\right\rangle
-w_{k}\right) p_{k}+w_{k}\pi _{k}$ and the value of its gradient $\nabla
\Phi _{\mu ,\pi }\left( w\right) =\pi -p$, ie. $\frac{\partial \Phi _{\mu
,\pi }\left( w\right) }{\partial w_{k}}=\pi _{k}-p_{k}$. We have implemented
these calculations in Matlab using a modified versions of the publicly
available Multi-Parametric Toolbox (MPT)\footnote{%
MPT is available online at http://control.ee.ethz.ch/~mpt/.}. All the
programs are available upon request.

\section*{Conclusion}

In comparison with existing literature on the topic on multidimensional risk
exposures, this work proposes a multivariate extension of the notion of
comonotonicity, which involves simultaneous optimal rearrangements of two
vectors of risk. With this extension, we are able to generalize Kusuoka's
result and characterize subadditive, comonotonic additive and law invariant
risk measures by maximal correlation functionals, which we show can be
conveniently computed using optimal transportation methods. We also show
that the properties of law invariance, subadditivity and comonotonic
additivity can be summarized by an equivalent property, that we call \emph{%
strong coherence}, and that we argue has a more natural economic
interpretation. Further, we believe that this paper illustrates the enormous
potential of the theory of optimal transportation in multivariate analysis
and higher dimensional probabilities. We do not doubt that this theory will
be included in the standard probabilistic toolbox in a near future.

\textit{$\dag $Canada Research Chair in Mathematical Economics, University
of British Columbia. E-mail: ekeland@math.ubc.ca}

\textit{$\S $Corresponding author. \'Ecole polytechnique, Department of
Economics, 91128 Palaiseau, France. E-mail: alfred.galichon@polytechnique.edu%
}

\textit{$\ddag $ D\'epartement de sciences \'economiques, Universit\'e de
Montr\'eal, CIRANO, CIREQ. E-mail: marc.henry@umontreal.ca}

\appendix

\section{Illustrations}

The t\^atonnement algorithm was implemented with the use of the
Multi-Parametric Toolbox, and we derived the general quantile $\nabla V$
that achieves the optimal transportation of the uniform distribution on the
unit cube in $\mathbb{R}^d$ and the empirical distribution of a sample of
uniformly distributed random vectors in the unit cube in $\mathbb{R}^d$. The
following illustrations show the Monge-Kantorovitch potential $V$, also
interpreted as the buyer's indirect utility in our general equilibrium
interpretation in the case of samples of size 7 and 27 respectively. The
potential $V$ is piecewise affine, and the algorithm also requires to
determine the regions over which it is affine, and their volume and center
of mass. The corresponding partition is given opposite each potential plot.
For illustration purposes, the dimension of the space $d$ is taken equal to
2, but the generalized quantiles and corresponding partitions can be derived
in higher dimensions.

\begin{figure}[h!]
    \centering
    \includegraphics[width=\textwidth]{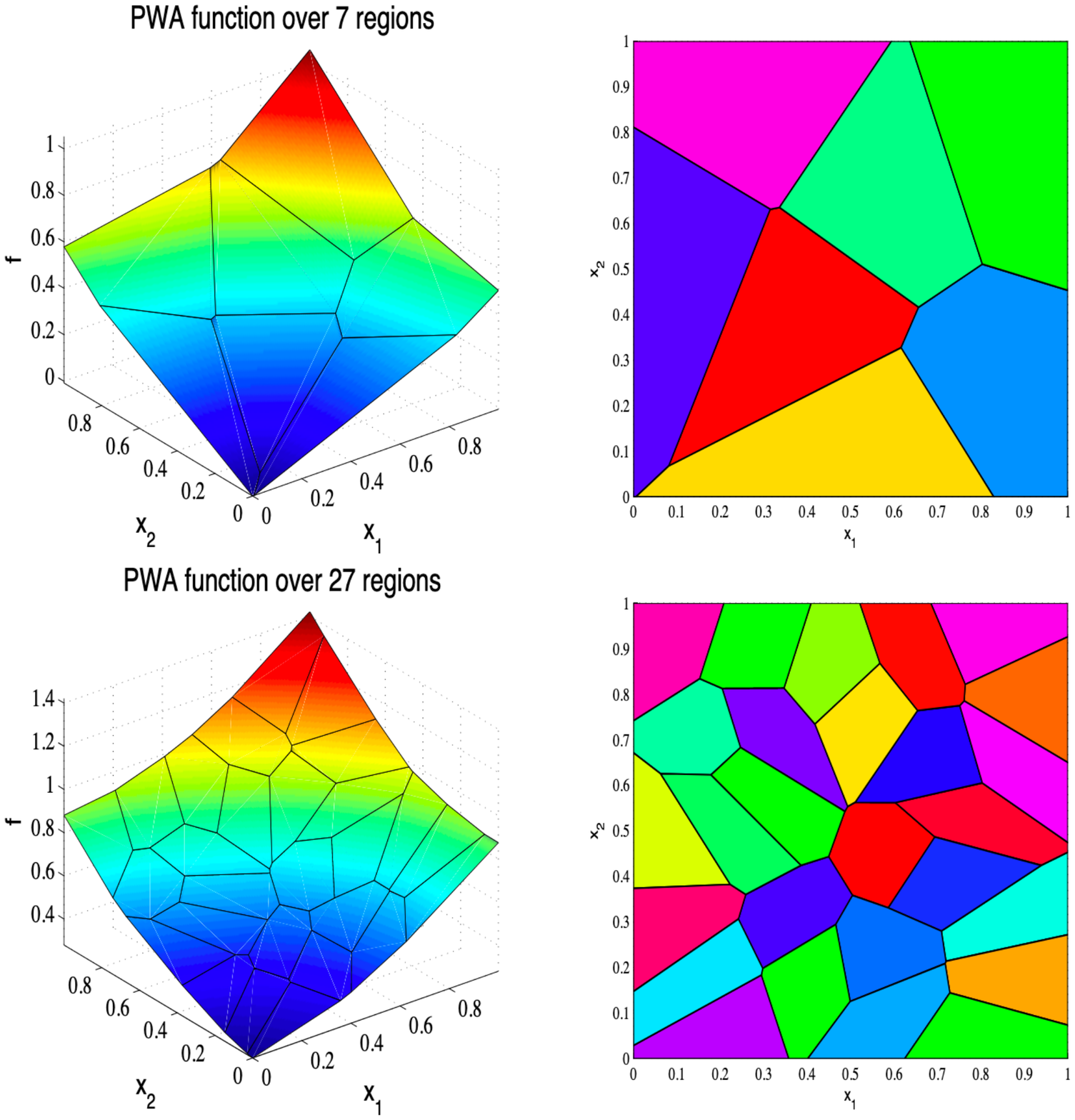}
    \caption{Mapping the uniform to a discrete discrete distribution in dimension $d=2$. Upper row: seven atom points, lower row: twenty-seven atom points. Left column: the potential $V(u)=w^*(u)$. Right column: the corresponding partition of the space $\mathcal{U}$.}
    \label{Figure:Discrete}
\end{figure}

\begin{figure}[h!]
    \centering
    \includegraphics[width=\textwidth]{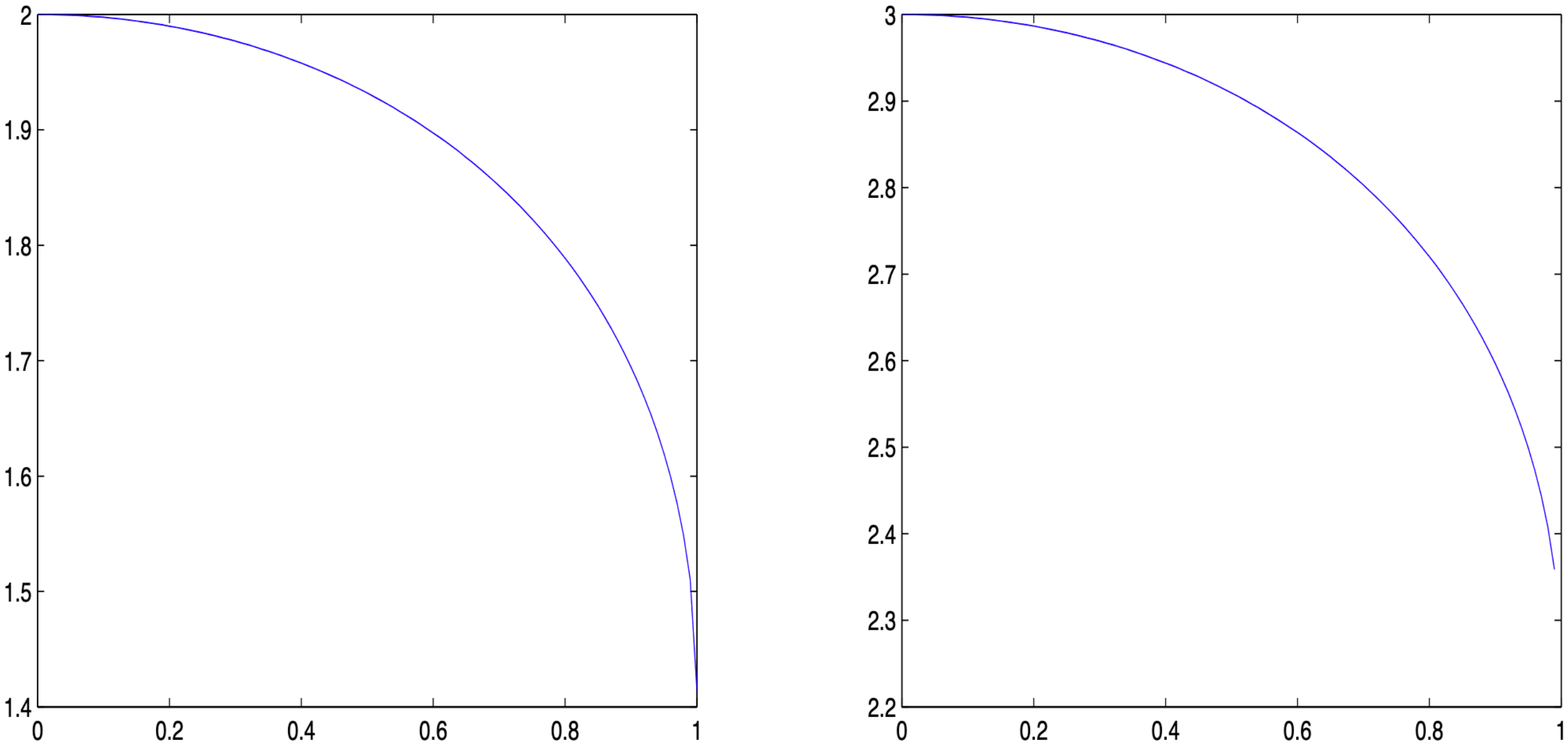}
    \caption{The value of the risk measure in the Gaussian case, plot against $\protect\rho$. Left: $\protect\sigma_1=\protect\sigma_2=1$. Right: $\protect\sigma_1=1, \protect\sigma_2=2$.}
    \label{Figure:Gauss}
\end{figure}

\section{Results on Optimal Transportation}

In this appendix we recall basic results in Optimal Transportation theory.
Roughly put, this theory characterizes the properties of the couplings of
two random variables which achieve maximal correlation. We state the
following basic result, due to Brenier (cf. \cite{Villani:2003}, Th. 2.12,
in which a proof is given).

\begin{proposition}
Let $\varrho$ be a maximal correlation risk measure with respect to baseline
risk $U$. Then if both $\mathcal{L}_U $ and $\mathcal{L}_X $ are absolutely
continuous, there exist a convex functions $V:{\mathbb{R}}^d\to{\mathbb{R}}$
and $W:{\mathbb{R}}^d\to{\mathbb{R}}$ which are Legendre-Fenchel conjugate
of each other ie. $W=V^*$, and
\begin{eqnarray*}
\varrho(X)&=&E[U\cdot\nabla V(U)] \\
\varrho(X)&=&E[X\cdot\nabla W(X)]
\end{eqnarray*}
where the map $\nabla V$ pushes forward $\mathcal{L}_U$ to $\mathcal{L}_X$,
and conversely $\nabla W$ pushes forward $\mathcal{L}_X$ to $\mathcal{L}_U$,
and $\nabla W= (\nabla V)^{-1}$. When only $\mathcal{L}_U $ is absolutely
continuous, then only those among the statements above involving $V$ alone
hold, and similarly, when only $\mathcal{L}_X $ is absolutely continuous
then only those among the statements above involving $W$ alone hold. \label%
{proposition: maxcorrel}
\end{proposition}

As $Q_X = \nabla V$ pushes forward measure $\mu$ on the distribution of $X$,
it can be seen in some sense as a natural extension of a univariate quantile
function (where $\mu=\mathcal{U}([0,1])$ - in which case $Q_X = F_X^{-1}$)
to the multivariate setting.

\end{document}